\documentclass[]{pasj02}
\usepackage{natbib}
\usepackage{bm} %追加,documentclassの後
\usepackage[switch,mathlines]{lineno} % add line number to manuscript
\jyear{2026}
\Received{}%{yyyy/mm/dd}
\Accepted{}%{yyyy/mm/dd}
\begin{document} 
\title{Beyond the $\alpha$ model: scaling the wind-driven accretion rate in protoplanetary disks using systematic non-ideal magnetohydrodynamical simulations}
%%% begin:list of authors
% Do NOT capitalize all letters in "textsc".
\author{
 Haruhi \textsc{Enomoto},\altaffilmark{1}\altemailmark\orcid{0000-0001-6390-8700} \email{haruhi.enomoto@gmail.com} 
 Shoji \textsc{Mori},\altaffilmark{2}\orcid{0000-0002-7002-939X}
 and 
 Satoshi \textsc{Okuzumi}\altaffilmark{1}\altemailmark\orcid{0000-0002-1886-0880} 
\email{okuzumi@eps.sci.isct.ac.jp} 
}
\altaffiltext{1}{Department of Earth and Planetary Sciences, Institute of Science Tokyo, 2-12-1 Ookayama, Meguro, Tokyo 152-8551, Japan}
\altaffiltext{2}{Institute for Advanced Study and Department of Astronomy, Tsinghua University, Beijing 100084,
People's Republic of China}
  %https://academic.oup.com/pasj/pages/Pasj_Keywords 
\maketitle
\begin{abstract}
Magnetically driven mass accretion plays a key role in protoplanetary disk evolution and planet formation. However, the $\alpha$ prescription remains phenomenological, and how the accretion rate depends on basic disk quantities is still poorly understood.
While local shearing-box simulations are computationally efficient, they suffer from a fundamental problem: the toroidal magnetic field generated by Keplerian shear accumulates within the computational domain, disrupting a field-line geometry consistent with global wind-driven accretion.
In this study, we use the super-box-scale diffusion (SBD) scheme in non-ideal MHD shearing-box simulations. By damping the horizontally averaged horizontal magnetic fields, this scheme successfully mitigates the artificial field accumulation and maintains the field-line symmetry required for global wind-driven accretion for more than 500 orbital periods.
Comparison with self-similar solutions supports the quantitative usefulness of the SBD method, showing good agreement in both the vertical structure and the plasma-beta dependence of the accretion rate.
We then conduct a parameter survey using a magnetic diffusivity table, covering a wide range of disk radii, surface densities, magnetic field strengths, and dust-to-gas ratios.
We demonstrate that the mass accretion rate follow power-law scaling relations in terms of three local disk properties: the midplane plasma beta, an effective ambipolar Elsasser number in ionized surface layers, and the thickness of the magnetically active layer.
The scaling relations reproduce the numerical results to within a factor of 2--3 across the explored parameter space.
The present scaling relations provide a framework for predicting the mass accretion rate from local disk physical quantities without invoking an $\alpha$ parameter.

\end{abstract}
%\pagewiselinenumbers 
\section{Introduction}
Accretion in protoplanetary disks plays a crucial role in determining the final masses and compositions of planets. Systematic surveys of nearby star-forming regions have been characterizing disk accretion rates across a wide range of stellar masses and ages \citep{2023ASPC..534..539M}.

However, significant uncertainties remain in current theoretical models of accretion disks. The classical viscous disk model \citep{1974MNRAS.168..603L} has been widely used as a phenomenological prescription of disk accretion. However, this model parameterizes the mass transport efficiency through a single dimensionless viscosity parameter $\alpha$ \citep{1973A&A....24..337S} whose connection to the underlying physical processes is ambiguous. In particular, it is widely assumed that $\alpha$ is spatially and temporally constant, but there is no guarantee that this assumption holds true \citep[e.g.,][]{2013ApJ...778L..14A,2017A&A...600A..75B,2017ApJ...845...75B,2022A&A...658A..97D,2024PASJ...76..616I}. Identifying the physical mechanism of mass transport is crucial for a better understanding of disk evolution.

The leading candidate for the mechanism driving mass transport in protoplanetary disks is angular momentum transport by magnetic fields (e.g., \citealt{1982MNRAS.199..883B,1989ApJ...342..208K,1991ApJ...376..214B,1995ApJ...440..742H}). 
Three-dimensional magnetohydrodynamics (MHD) simulations (e.g., \citealt{2013ApJ...769...76B,2014A&A...566A..56L,2015ApJ...801...84G,2021A&A...650A..35L,2024PASJ...76..616I}) that incorporate non-ideal MHD effects (e.g., \citealt{1996ApJ...457..355G,2000ApJ...543..486S,2007Ap&SS.311...35W,2014prpl.conf..411T,2023ASPC..534..465L}) have shown that magnetic winds serve as the dominant angular momentum transport mechanism in weakly ionized regions where magnetic turbulence is suppressed (for a review, see, e.g., \citealt{2023ASPC..534..465L}). 
In this picture, angular momentum is removed vertically by magnetically driven outflows launched from the disk surface, thereby allowing the disk gas to accrete.
The magnetically driven mass accretion rate in such regions depends strongly on both the ionization fraction and the net vertical magnetic field strength (e.g., \citealt{2013ApJ...775...73S,2017A&A...600A..75B,2017ApJ...845...75B,2020ApJ...896..126G,2021A&A...650A..35L,2021MNRAS.507.1106C}). 
However, no empirical model yet exists that describes how the accretion rate depends on all these disk parameters. Developing such a model requires systematic MHD simulations of disk accretion under realistic ionization fraction distributions.

Previous MHD simulations of protoplanetary disk accretion can be classified into radially global and local approaches. The radially global approach \citep[e.g.,][]{2016A&A...596A..74S,2017A&A...600A..75B,2018MNRAS.477.1239S,2019MNRAS.484..107S,2019PASJ...71..100S,2020ApJ...896..126G,2023ApJ...957...99S} has the advantage of naturally treating global magnetic field-line geometry and radial magnetic flux transport. However, this approach generally requires high computational costs and is therefore not well-suited for extensive parameter surveys. Recently,  \citet{2021A&A...650A..35L} proposed a self-similar approach that reduces the global problem to a one-dimensional one by assuming radial self-similarity. This method can capture global accretion properties at low computational cost; however, it is not applicable to realistic ionization fraction distributions, which are not necessarily self-similar in radius.

The radially local approach \citep[e.g.][]{1995ApJ...440..742H,1996ApJ...463..656S,2009ApJ...691L..49S,2011ApJ...742...65O,2013ApJ...769...76B,2015MNRAS.454.1117S,2018A&A...617A.117R} employs a shearing box, in which the dynamical equations in a local Cartesian domain centered at a given orbital radius are solved in the presence of Keplerian shear.
Local simulations are a powerful tool for computing the vertical structure at any given radius with high resolution and low computational cost.
However, because the shearing box assumes uniform shear and periodic boundary conditions, it does not fully capture the radial gradient of the toroidal field produced by Keplerian rotation, nor the associated radial diffusion. As a result, the toroidal magnetic flux generated by shear tends to accumulate within the local domain (e.g., \citealt{2008ApJ...679L.131T}).
This excessive accumulation of toroidal magnetic flux makes it difficult to maintain a field-line configuration consistent with the system's global geometry, resulting in an unphysical accretion structure that is perfectly anti-symmetric with respect to the disk midplane, thereby yielding zero net accretion \citep[e.g.,][]{2013ApJ...769...76B,2014A&A...566A..56L}.

%%%\red{5.本研究の目的}%%%%%%%%%%%%%%%%%%%%%%%%%%%%
In this study, we present a systematic study of magnetically driven accretion in weakly ionized protoplanetary disks by using a refined local shearing-box approach that mitigates the accumulation of horizontal magnetic fields. Our method employs the super-box-scale diffusion (SBD) scheme \citep{2012MNRAS.422.1140G}, which regularly removes the horizontally averaged component of the horizontal magnetic fields from the local box. We demonstrate that this SBD scheme enables us to compute vertical profiles of gas accretion that are consistent with global self-similar solutions \citep{2021A&A...650A..35L}. Using this approach, we derive an empirical law for the magnetically driven accretion rate, including the effects of Ohmic and ambipolar diffusion, as a function of the disk's vertical ionization structure and the net vertical magnetic flux. 

%%%\red{6.本論文の構成}%%%%%%%%%%%%%%%%%%%%%%%%%%%%
This paper is organized as follows. In section \ref{sec:method}, we describe the numerical methods and the formulation of the SBD scheme. In section \ref{sec:Validation}, we examine the numerical stability of the SBD scheme and then validate the scheme against self-similar solutions. In section \ref{sec:Scaling}, we present the results of an extensive parameter survey using a magnetic diffusivity table based on ionization equilibrium calculations, systematically investigate the effects of disk radius, surface density, and dust-to-gas ratio on the accretion rate, and propose the resulting scaling laws. Finally, in section \ref{sec:discussion}, we discuss implications of our findings and future directions.

\section{Method}\label{sec:method}
In this study, we numerically solve the conservative non-ideal MHD equations in the local shearing-box approximation (subsection~\ref{ssec:method_equation}).
To mimic the global diffusive relaxation of the mean magnetic field, we add an SBD term to the induction equation (subsection~\ref{ssec:method_SBD}).
We adopt two prescriptions for the magnetic diffusivities: an analytic parametric model and a tabulated model based on ionization equilibrium calculations (subsection~\ref{sec:diffusivities}). Subsection~\ref{ssec:method_numerical_scheme} summarizes the numerical scheme and run parameters, and subsection~\ref{ssec:method_accretion_rate} defines the mass accretion rate and explains our treatment of the field-line geometry.

\subsection{Governing equations}\label{ssec:method_equation}
We introduce a local Cartesian coordinate system $(x, y, z) = (r - r_0,\, r_0(\phi - \Omega_{\rm K} t),\, z)$ centered at an orbital radius $r_0$ and corotating with the local Keplerian angular velocity $\Omega_{\rm K}$. Here, $r$ is the cylindrical radius, $\phi$ is the azimuthal angle, $z$ is the vertical distance from the disk midplane, and $t$ is time.
Throughout this paper, we identify the $x$- and $y$- directions with the $r$- and $\phi$- directions, respectively, unless otherwise noted. For example, the radial component of the magnetic field is denoted by either $B_x$ or $B_r$, depending on the context. For clarity, we hereafter omit the subscript 0 from $r_0$.

The continuity equation is
\begin{equation}
  \frac{\partial \rho}{\partial t} + \nabla \cdot (\rho \bm{v}) = 0,
\end{equation}
where $\rho$ is the gas density.
The momentum equation is
\begin{equation}
\begin{aligned}
  \frac{\partial (\rho \bm{v})}{\partial t}
  + \nabla \cdot \left[ \rho \bm{v}\bm{v}
  - \frac{\bm{B}\bm{B}}{4\pi}
  + \left( P + \frac{B^2}{8\pi} \right) \bm{I} \right]=\\
  - 2 \rho {\Omega}_{\rm K}\hat{\bm z} \times \bm{v}
  + 3 \rho \Omega_{\rm K}^2 x \, \hat{\bm{x}}-\rho\,\Omega_{\rm K}^2 z\,\hat{\boldsymbol{z}},
\end{aligned}
\end{equation}
where $\bm{v}$ is the gas velocity vector, $\bm{B}$ is the magnetic field vector, $P$ is the gas pressure, $\bm{I}$ is the unit tensor, and $\hat{\bm{x}}$ and $\hat{\bm{z}}$ are the unit vectors in the $x$- and $z$-directions, respectively.

We adopt the local isothermal approximation and relate $P$ to $\rho$ as
\begin{equation}
  P = c_{\rm s}^2 \rho, \label{eq:method_isothermal}
\end{equation}
where $c_{\rm s}$ is the isothermal sound speed. The isothermal scale height is given by $H \equiv c_{\rm s}/\Omega_{\rm K}$.

We include Ohmic and ambipolar diffusion, while neglecting the Hall effect. The magnetic field evolves according to the induction equation,
\begin{equation}
  \frac{\partial \bm{B}}{\partial t}
  = \nabla \times \left( \bm{v} \times \bm{B}
  - \eta_{\rm O} \bm{J}
  - \eta_{\rm A} \bm{J}_\perp \right)
  + \left. \frac{\partial \bm{B}}{\partial t} \right|_{\rm SBD}, \label{eq:induction_with_SBD}
\end{equation}
where the current density $\bm{J}$ is defined by Ampère's law,
\begin{equation}
  \bm{J} = \frac{c}{4\pi} \nabla \times \bm{B}.
\end{equation}
The last term, $\left.\partial \bm{B}/\partial t\right|_{\rm SBD}$, represents the SBD contribution, which damps the horizontally averaged magnetic field. Its explicit form is given in subsection~\ref{ssec:method_SBD}.

The initial density profile is given by the Gaussian profile of an isothermal hydrostatic atmosphere,
\begin{equation}
    \rho(z) = \rho_0 \exp\!\left( -\frac{z^2}{2H^2} \right), \label{eq:def_rho_exp}
\end{equation}
where $\rho_0$ is the initial density at the midplane ($z = 0$).
We initialize the magnetic field with a spatially uniform vertical component $B_{z0}$.
Defining the initial plasma beta as $\beta_0$, the isothermal equation of state (equation~\eqref{eq:method_isothermal}) gives
\begin{equation}
\beta_0 \equiv \frac{8\pi \rho_0 c_s^2}{B_{z0}^2},  \label{eq:beta0_definition}
\end{equation}
and the plasma beta $\beta$ is defined as
\begin{equation}
\beta \equiv \frac{8\pi P}{B^2}. 
\end{equation}

\subsection{SBD prescription}\label{ssec:method_SBD}
The local shearing-box approximation cannot treat global radial diffusion of magnetic flux self-consistently. As a result, the toroidal magnetic field persistently accumulates within the shearing box and can eventually lead to unphysical field configurations \citep[see, e.g.,][]{2013ApJ...769...76B,2014A&A...566A..56L}.
In reality, any magnetic field that accumulates locally can diffuse away over a global diffusion timescale. The SBD prescription employed by \citet{2012MNRAS.422.1140G} enables us to mimic this large-scale relaxation in local simulations.

The SBD prescription decays the horizontally averaged horizontal magnetic fields at a constant rate as
\begin{equation}
\left.\frac{\partial B_i}{\partial t}\right|_{\rm SBD}
=-\frac{\langle B_i\rangle_{xy}}{\tau_{\rm diff}(z)}
\qquad (i=x,y),
\end{equation}
where the angle brackets $\langle \cdots \rangle_{xy}$ denote a horizontal average and 
$\tau_{\rm diff}(z)$ is the global magnetic diffusion timescale. Here the SBD term is applied only for \(i=x,y\), while 
\(\left.\partial B_z/\partial t\right|_{\rm SBD}=0\). Approximating the global variation scale of the magnetic field as $\sim r$ and defining the effective magnetic diffusivity as $\bar{\eta}\equiv \eta_{\rm A} + \eta_{\rm O}$, the global diffusion timescale can be estimated as $\tau_{\rm eff} \sim r^2/\bar{\eta}$. We therefore take $\tau_{\rm diff}$ to be
\begin{equation}
\tau_{\rm diff}(z)=\frac{r^2}{C \langle \bar{\eta}\rangle_{xy}},
\label{eq:SBD_tau_diff}
\end{equation}
where $C$ is a dimensionless coefficient. 
The corresponding SBD term can then be rewritten as
\begin{equation}
\left.\frac{\partial B_i}{\partial t}\right|_{\rm SBD}
=-\frac{C \langle \bar{\eta}\rangle_{xy}}{r^2}
\langle B_i\rangle_{xy}
\qquad (i=x,y).
\label{eq:SBD}
\end{equation}
In the numerical implementation, $C$ serves as a parameter that controls the efficiency of the global field relaxation. Unless otherwise stated, we adopt $C = \pi^2$ throughout this paper. 
This choice corresponds to approximating the large-scale spatial gradient of the magnetic field by the wavelength of the longest Fourier mode (equation~(B3) of \citet{2012MNRAS.422.1140G}).
The SBD prescription does not violate the divergence-free condition $\nabla\!\cdot\!\bm{B}=0$ as it only subtracts the uniform components of the horizontal fields.

\subsection{Magnetic diffusivities}\label{sec:diffusivities}
Ohmic diffusion and ambipolar diffusion are characterized by the diffusivities $\eta_{\rm O}$ and $\eta_{\rm A}$, respectively. The relative importance of magnetic diffusion compared to magnetic induction can be quantified by the Ohmic and ambipolar Elsasser numbers,
\begin{equation}
\Lambda_{\rm O}\equiv \frac{v_{\rm A}^2}{\eta_{\rm O}\Omega_{\rm K}}, \label{eq:def_LambdaO}
\end{equation}
\begin{equation}
{\rm Am}\equiv \frac{v_{\rm A}^2}{\eta_{\rm A}\Omega_{\rm K}},
\label{eq:def_Am}
\end{equation}
where $v_{\rm A}=|\bm{B}|/\sqrt{4\pi\rho}$ is the Alfv\'{e}n speed. Ohmic and ambipolar diffusion become important when $\Lambda_{\rm O}\lesssim 1$ and ${\rm Am}\lesssim 1$, respectively
(e.g., \citealt{2000ApJ...543..486S}; \citealt{2011ApJ...736..144B}).
The magnetic Reynolds number,
\begin{equation}
{\rm Rm} \equiv \frac{c_{\rm s}^2}{\eta_{\rm O}\Omega_{\rm K}}
= \left(\frac{c_{\rm s}}{v_{\rm A}}\right)^2 \Lambda_{\rm O},
\end{equation}
also serves as a useful diagnostic for Ohmic diffusion (e.g., \citealt{2000ApJ...530..464F}).

In this study, we employ two prescriptions for the magnetic diffusivities: an analytic parametric model used by \citet{2021A&A...650A..35L} and a tabulated model based on ionization equilibrium calculations.
These prescriptions are described below.

\subsubsection{Parametric diffusivity model}\label{subsubsec:paramdiffmodel}
\begin{figure}[t]
 \begin{center}
  \includegraphics[width=\columnwidth]{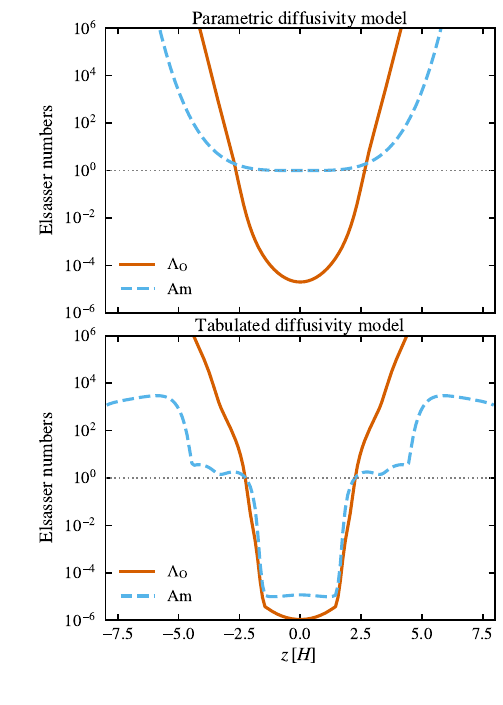}
 \end{center}
\caption{
Vertical profiles of the Elsasser numbers $\Lambda_{\rm O}$ (solid) and ${\rm Am}$ (dashed) as a function of height $z/H$.
The dotted line marks ${\rm Rm}, {\rm Am} = 1$. 
and the lower panel shows the result for the tabulated diffusivity model
(subsubsection~\ref{subsubsec:tabdiffmodel}) with $r = 1~{\rm au}$, $\Sigma = 10^3~{\rm g~cm^{-2}}$, $\beta_0 = 10^4$, $f_{\rm dg} = 10^{-5}$, and $T = 110~{\rm K}$.
Both panels use the same axis scales. Alt text: Two line graphs arranged vertically, sharing the same logarithmic vertical axis from 10 to the minus 6 to 10 to the 6 and horizontal axis from minus 8 to 8 scale height. In each panel, the solid line traces the Ohmic Elsasser number, and the dashed line traces the ambipolar Elsasser number Am as functions of height.}
\label{fig:LambdaO_Am_two_panels}
\end{figure}

In the parametric model, we prescribe ${\rm Rm}$ and ${\rm Am}$ as functions of $z$, following \citet{2021A&A...650A..35L}:
\begin{equation}
\begin{aligned}
  {\rm Rm}(z) &= {\rm Rm}_0 \exp\!\left[\left(\frac{z}{3H}\right)^4\right]
  \left(\frac{\rho(z)}{\rho(0)}\right)^{-1}, \\
  {\rm Am}(z) &= {\rm Am}_0 \exp\!\left[\left(\frac{z}{3H}\right)^4\right]
\end{aligned}
\label{eq:Am_profile}
\end{equation}
where ${\rm Rm}_0$ and ${\rm Am}_0$ are the midplane values.
The corresponding diffusivities $\eta_{\rm O} = {v_{\rm A}^2}/({\Lambda_{\rm O}\Omega_{\rm K}})$ and $\eta_{\rm A} = {v_{\rm A}^2}/({{\rm Am}\Omega_{\rm K}})$ increase gradually from the midplane toward the disk surface, mimicking the resistivity structure in a protoplanetary disk with ionized surfaces.
The upper panel of figure~\ref{fig:LambdaO_Am_two_panels} shows the vertical profiles of $\Lambda_{\rm O}$ and ${\rm Am} $ for $(\beta_0, {\rm Rm}_0, {\rm Am}_0) = (10^5, 1, 1)$, where the density profile $\rho(z)$ follows equation~\eqref{eq:def_rho_exp}.

\subsubsection{Tabulated diffusivity model} \label{subsubsec:tabdiffmodel}
The tabulated diffusivity model uses the magnetic diffusivity table adopted by \citet{2019ApJ...872...98M}. This table is constructed based on ionization equilibrium calculations (see subsection~2.3 of \citealt{2016ApJ...817...52M}) that account for ionization by cosmic rays, stellar X-rays, and radioactive nuclide decay.
Assuming local ionization equilibrium, we compute the number densities of charged particles, including dust-grain charge states, and then evaluate $\eta_{\rm O}$ and $\eta_{\rm A}$.
The table is constructed for a fixed dust grain size of $0.1\,\mu{\rm m}$.
At fixed grain size, varying the dust-to-gas mass ratio $f_{\rm dg}$ approximately changes the total dust surface area, which primarily controls the recombination rate in ionization equilibrium.
Following \citet{2019ApJ...872...98M}, we assume low small-grain abundances of $f_{\rm dg}=10^{-5}$--$10^{-3}$ rather than the interstellar dust-to-gas ratio of $\sim 0.01$. This choice represents disks in which most submicron grains have grown into larger solids or settled toward the midplane, reducing the grain surface area available for recombination.  Depletion of small grains by a factor of 100--1000 is consistent with constraints from disk infrared observations \citep[e.g.,][]{2005ApJ...628L..65F, 2026arXiv260623815B}.
For each simulation run, we specify the dust-to-gas mass ratio $f_{\rm dg}$ and the gas temperature $T$, and use the corresponding table to evaluate the diffusivity coefficients. We also include the ionization of C and S by FUV radiation in addition to the tabulated magnetic diffusivities (\citealt{2011ApJ...735....8P}; see \citealt{2013ApJ...769...76B} for details).
To avoid excessively small timesteps due to strong magnetic diffusion, we impose an upper limit of $100 c_{\rm s} H$ on both $\eta_{\rm O}$ and $\eta_{\rm A}$. This diffusivity cap is not applied to the SBD term. We confirmed that even increasing the cap value by a factor of 10 does not significantly change the quasi-steady structure or the mass accretion rate.

Using the tabulated diffusivity model, we run a total of 46 simulations covering $r=\{1,3,10,30\}\,{\rm au}$, $\Sigma=\{10^2,10^3,10^4\}\,{\rm g~cm^{-2}}$, $\beta_0=\{3.2\times10^3,10^4,3.2\times10^4,10^5,3.2\times10^5,10^6\}$, and $f_{\rm dg}=\{10^{-5},10^{-4},10^{-3}\}$.
The input parameters for these runs are listed in table~\ref{tab:run_list_main} (appendix~\ref{appndix:Data}).
The lower panel of figure~\ref{fig:LambdaO_Am_two_panels} shows the vertical profiles of $\Lambda_{\rm O}$ and ${\rm Am}$ from the tabulated diffusivity model with $r=1~{\rm au}$, $\Sigma=10^3~{\rm g~cm^{-2}}$, $\beta_0=10^4$, $f_{\rm dg}=10^{-5}$, and $T=110\,{\rm K}$.
The disk structure resulting from our tabulated diffusivity model typically consists of an Ohmic dead zone with $\Lambda_{\rm O} <1$, an intermediate magnetically active layer with ${\rm Am}\sim O(1)$, and a wind region with $\beta \lesssim 1$. Although the detailed vertical profiles depend on the adopted ionization chemistry and dust models, qualitatively similar transitions from an Ohmic dead zone to an ambipolar-dominated surface layer appear in previous non-ideal MHD disk models \citep[e.g.,][]{2013ApJ...769...76B,2017A&A...600A..75B,2017ApJ...845...75B}. We therefore regard this table as a representative layered ionization structure of weakly ionized protoplanetary disks.

\subsection{Numerical scheme and run parameters}\label{ssec:method_numerical_scheme}
We solve the non-ideal MHD equations in the local shearing-box approximation
using Athena \citep{2008ApJS..178..137S}.
We compute MHD fluxes with the Harten--Lax--van Leer--discontinuities
Riemann solver and use a second-order Godunov scheme.
The constrained transport method maintains the divergence-free constraint $\nabla\!\cdot\!\bm{B}=0$.
We use a second-order Runge--Kutta scheme for time integration, with the Courant--Friedrichs--Lewy number set to $0.3$.
The diffusion terms are updated using super-time-stepping for numerical
efficiency and stability in regions with high diffusivities.

The computational domain covers $x \in [-0.5,\,0.5]H$, $y \in [-1,\,1]H$, and $z \in [-8,\,8]H$, with a uniform grid of $(N_x,\,N_y,\,N_z) = (16,\,16,\,512)$ cells.
We apply shearing-periodic boundary conditions in the $x$-direction, periodic boundary conditions in the $y$-direction, and outflow boundary conditions in the $z$-direction, where inflow is prohibited.

We implement several numerical treatments and verification procedures to ensure the stability and robustness of our simulations.
We add small random perturbations to the initial density and velocity fields to seed instabilities.
To avoid numerical instabilities caused by excessive magnetization in low-density regions, we impose a density floor determined by $\beta_0$ at every time step, setting the density to $\rho_{\rm floor}=0.01\beta_0^{-1}$ in units of the initial midplane density whenever the calculated density falls below this value.
This floor serves to stabilize the calculation during the initial transient phase, and we confirmed that it is not reached after each run has entered a quasi-steady state.
To confirm that our results are insensitive to the horizontal domain sizes, we have rerun the fiducial model with a wider horizontal domain of $x \in [-2,\,2]H$ and $y \in [-4,\,4]H$. The result shows that the horizontally averaged profiles of the density, magnetic field, and velocities agree with those of the fiducial run to within $\sim0.7\%$ accuracy.
Finally, each simulation is run for 500 orbital periods to confirm that the disk’s vertical structure reaches a quasi-steady state.

We adopt $c_s$, $\Omega_{\rm K}^{-1}$, and $\rho_0$ as the units of velocity, time, and density, respectively.
The unit of length is the scale height $H$, and the unit of magnetic field is $B_{\rm u} \equiv \sqrt{4\pi\rho_0}\,c_s$.
Hereafter, we explicitly indicate the units for clarity.

\subsection{Accretion rate and field-line geometry}
\label{ssec:method_accretion_rate}
The goal of this study is to model the mass accretion rate. 
Following equation~(15) of \citet{2007Ap&SS.311...35W}, we define the mass accretion rate at cylindrical radius $r$ as
\begin{equation}
\dot{M} \equiv -2\pi r \int_{-z_{\rm surf}}^{+z_{\rm surf}} \rho v_r\,dz,
\label{eq:mdot_def}
\end{equation}
where $z_{\rm surf}$ denotes the height of the disk surface, defined by the condition that the magnetic pressure equals the gas pressure, i.e.,
\begin{equation}
\beta(z_{\rm surf}) = 1.  \label{eq:def_z_surf}
\end{equation}
Under the assumptions of axisymmetry and steady state, vertical integration of the angular momentum equation gives $\dot{M}$ in terms of the magnetic field (equation~(17) of \citealt{2007Ap&SS.311...35W}):
\begin{equation}
\dot{M}\approx
\frac{1}{r\Omega_K}\frac{\partial}{\partial r}\left[r^2 \int_{-z_{\rm surf}}^{+z_{\rm surf}}\left(-B_r B_\phi\right)\,dz\right]-
\frac{r}{\Omega_K}\left[B_z B_\phi\right]_{-z_{\rm surf}}^{+z_{\rm surf}}.
\label{eq:mdot_stress_general}
\end{equation}
On the right-hand side of the above equation, the first and second terms represent the contributions from radial angular momentum transport by the magnetic stress within the disk and from vertical angular momentum transport by the Maxwell stress $\pm(B_z B_\phi/4\pi)_{\pm z_{\rm surf}}$ acting on the upper and lower surfaces, respectively.

In wind-driven accretion, magnetic stresses acting on the upper and lower disk surfaces extract angular momentum from the disk.
The second term on the right-hand side of equation~\eqref{eq:mdot_stress_general} represents this net vertical extraction of angular momentum and drives accretion.
Extracting angular momentum from both sides in the same rotational sense requires $B_\phi$ to have opposite signs at the two surfaces assuming $B_z$ retains the same sign throughout the disk.
This configuration, where $B_\phi$ reverses sign at some height within the disk interior, has been confirmed in global simulations and self-similar solutions (e.g.,~\citealt{1982MNRAS.199..883B, 2021A&A...650A..35L}), and we refer to it as the \textit{physical field-line geometry} \citep[see also their figure~9]{2013ApJ...769...76B}.
In contrast, when $B_\phi$ has even symmetry about the midplane, the surface stresses from the upper and lower sides cancel each other in equation~\eqref{eq:mdot_stress_general}, and no net vertical angular momentum transport arises.
We refer to this configuration as the \textit{unphysical field-line geometry}.
By omitting the direction toward the central star and neglecting curvature terms, the standard shearing-box approximation loses the global geometric constraint that field lines should incline in the same radial direction on both sides of the disk.
As a result, both the physical and unphysical field-line geometries can arise as numerical solutions.
Indeed, \citet{2013ApJ...769...76B} reported that both configurations can appear in local simulations, and that the physical field-line geometry is difficult to maintain for long periods in the shearing box, so that the unphysical field-line geometry tends to be selected.
In the next section, we show that the SBD scheme stabilizes the physical field-line geometry.

\section{Validation of the SBD prescription}\label{sec:Validation}
In this section, we validate the SBD scheme using the parametric diffusivity model (section~\ref{subsubsec:paramdiffmodel}), which is the same prescription adopted by \citet{2021A&A...650A..35L} and thus enables a direct comparison with their self-similar solutions. 
In subsection~\ref{ssec:Validation_Time_evolution}, we demonstrate that SBD maintains the physical field-line geometry over long periods. 
In subsection~\ref{ssec:Validation_Compare_with_Lesur}, we compare the resulting vertical structure and mass accretion rate with the self-similar solutions.

In this and the following sections, all field quantities represent horizontal averages. For clarity, we omit the angle brackets denoting the horizontal average, $\langle \cdots \rangle_{xy}$, unless otherwise noted.

\subsection{Time evolution with and without SBD} \label{ssec:Validation_Time_evolution}

\begin{figure}
  \includegraphics[width=\columnwidth]{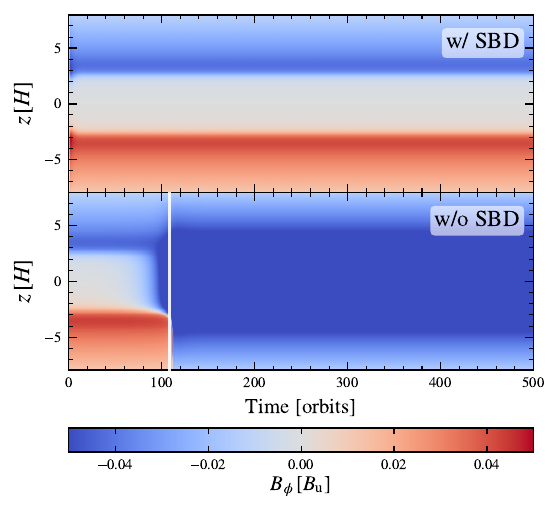} 
  \caption{
  Space--time diagram of the horizontally averaged toroidal field $B_\phi$ for the parametric diffusivity model with $(\beta_0, {\rm Rm}_0, {\rm Am}_0) = (10^5, 1, 1)$. The upper and lower panels show the results with and without SBD, respectively. The vertical white lines mark the time at which the field-line geometry transitions from the physical to the unphysical configuration. Alt text: Two two-dimensional color maps arranged vertically. The horizontal axis shows time from 0 to 500 orbits and the vertical axis shows height from minus 8 to 8 scale heights. The color indicates the toroidal field ranging from minus 0.05 to 0.05 in units of Bu.}
\label{fig:SBD_space-time_diagram}
\end{figure}

\begin{figure}[t]
  \includegraphics[width=\columnwidth]{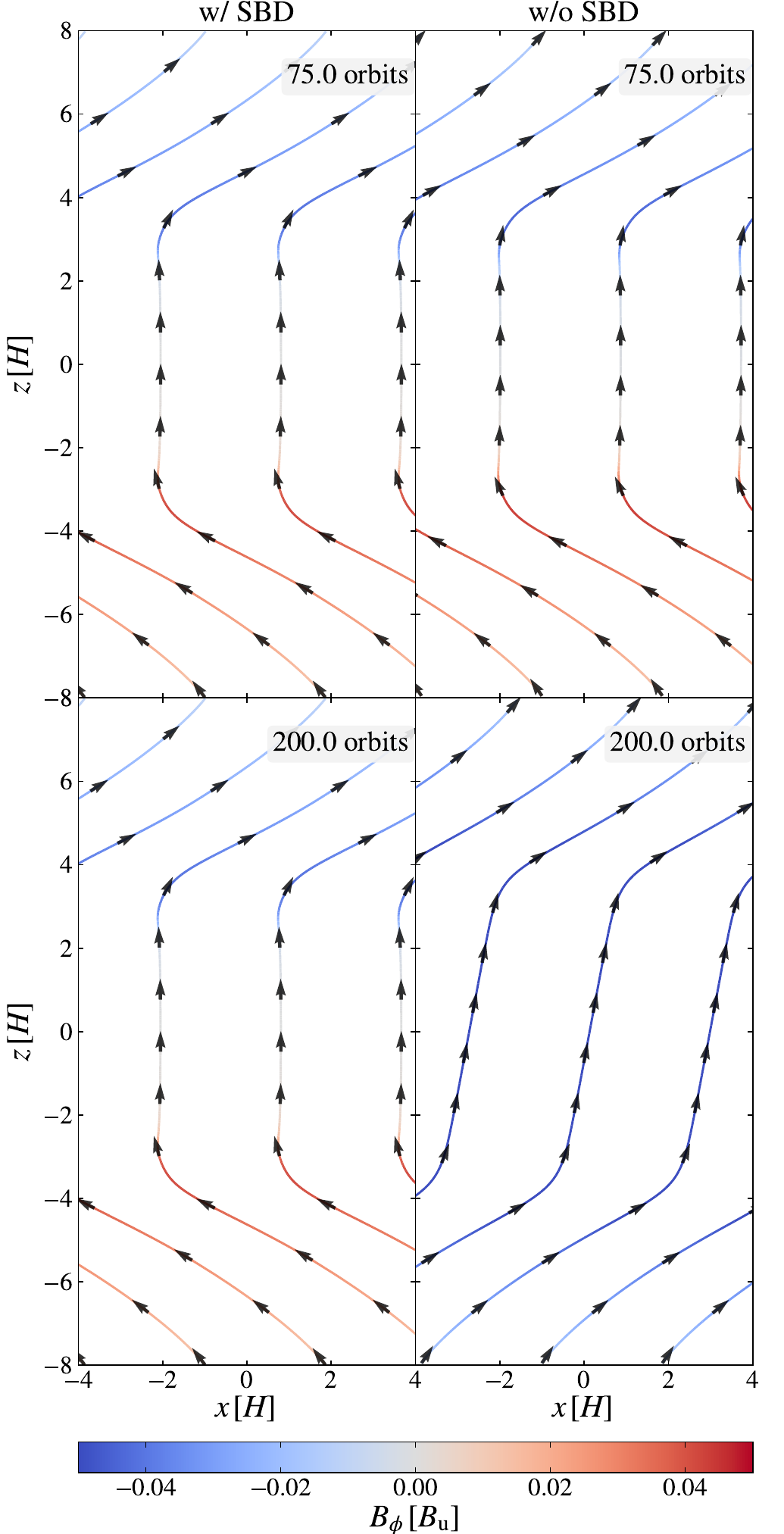}
  \caption{
  Magnetic field structure at 75 and 200 orbits for the parametric diffusivity model with $(\beta_0, {\rm Rm}_0, {\rm Am}_0) = (10^5, 1, 1)$. Streamlines show the poloidal field lines of $(B_x, B_z)$, and the color indicates the toroidal field $B_\phi$. Although the radial extent of the computational domain is $-0.5H < x < 0.5H$, the figure displays the range $-4H < x < 4H$ by assuming periodicity in the $x$-direction. The left and right columns show the results with and without SBD, respectively. Alt text: Four two-dimensional color maps in a two-by-two grid. The vertical axis ranges from minus 8 to 8 scale heights. The color indicates the toroidal field ranging from minus 0.05 to 0.05 in units of Bu.}
\label{fig:beta_75orbit_200orbit}
\end{figure}

We begin by illustrating the effect of the SBD on the long-term stability of the field-line geometry using the case with $(\beta_0, {\rm Rm}_0, {\rm Am}_0) = (10^5, 1, 1)$. We run two simulations for this case: one with SBD and one without. Both runs start from identical initial conditions, with only a uniform vertical magnetic field $B_z$ imposed. 

Figure~\ref{fig:SBD_space-time_diagram} compares the evolution of the vertical distribution of the azimuthal magnetic field $B_\phi$ in simulations with and without SBD. Both simulations initially develop a physical field-line geometry in which the sign of $B_\phi$ flips across the midplane. However, in the absence of SBD, this flip breaks at $t \approx 110$ orbits, resulting in an unphysical field-line configuration with no field-line reversal.
The right panels of figure~\ref{fig:beta_75orbit_200orbit} show snapshots of the field-line configuration before and after the transition of the field-line configuration. 

A closer inspection of the SBD-free run reveals that $B_\phi$ near the midplane grows in the negative direction until the physical field-line configuration breaks down. The background shear generates this toroidal field and continuously accumulates within the computational domain under periodic boundary conditions. At $t \approx 110$ orbits, the amplified toroidal field becomes strong enough to disrupt the field-line reversal around the midplane, likely due to magnetic tension.

In contrast, the run with SBD maintains the physical field-line geometry for at least 500 orbits (see figures~\ref{fig:SBD_space-time_diagram} and \ref{fig:beta_75orbit_200orbit}), as SBD suppresses the buildup of toroidal magnetic fields.

\begin{figure}[t]
  \includegraphics[width=\columnwidth]{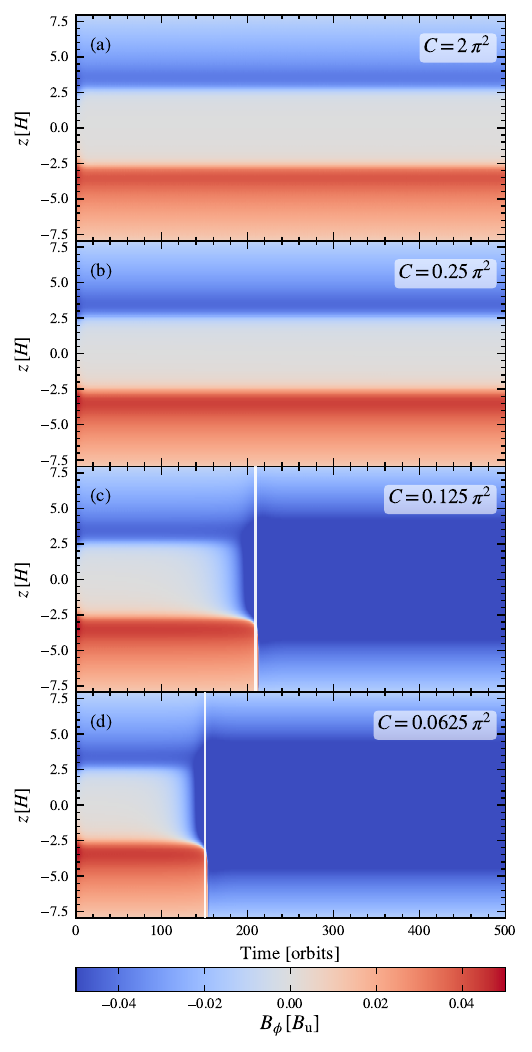}
  \caption{
  Space--time diagram of the horizontally averaged toroidal field $B_\phi$ for the parametric diffusivity model with $(\beta_0, {\rm Rm}_0, {\rm Am}_0) = (10^5, 1, 1)$. Panels (a) through (d) show the results for (a) $C = 2\pi^2$, (b) $0.25\pi^2$, (c) $0.125\pi^2$, and (d) $0.0625\pi^2$, from top to bottom. Alt text: Four two-dimensional color maps arranged vertically. The horizontal axis shows time from 0 to 500 orbits and the vertical axis shows height from minus 8 to 8 scale heights. The color indicates the toroidal field ranging from minus 0.05 to 0.05 in units of Bu.}
  \label{fig:SBD_param_C}
\end{figure}

\begin{figure}
  \includegraphics[width=\columnwidth]{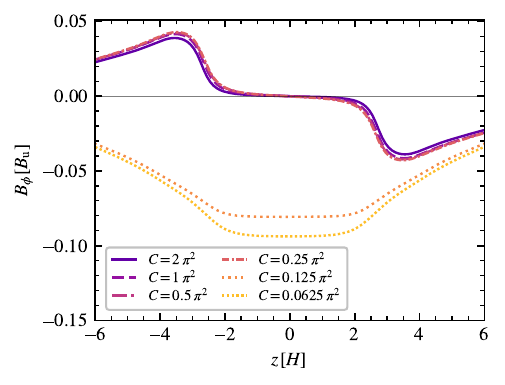}
  \caption{
  Horizontally and time-averaged vertical profiles of $B_\phi$ over 480--500 orbits from runs with different values of $C$, using the parametric diffusivity model with $(\beta_0, {\rm Rm}_0, {\rm Am}_0) = (10^5, 1, 1)$. Alt text: A line graph with six lines. The horizontal axis shows height from minus 6 to 6 scale heights and the vertical axis shows the toroidal field from minus 0.15 to 0.05 in units of Bu. The six lines correspond to C values of 2 pi squared, pi squared, 0.5 pi squared, 0.25 pi squared, 0.125 pi squared, and 0.0625 pi squared.}
  \label{fig:C_Bphi_param}
\end{figure}

We next examine how the stability of the physical field-line geometry depends on the dimensionless coefficient $C$ that controls the suppression of horizontal fields (see equation~\eqref{eq:SBD}). 
Figure~\ref{fig:SBD_param_C} shows the time evolution of $B_\phi$ for various values of $C$ in the model with $(\beta_0, {\rm Rm}_0, {\rm Am}_0) = (10^5, 1, 1)$. 
Figure~\ref{fig:C_Bphi_param} shows the time-averaged vertical profile of $B_\phi$ over 480--500 orbits.
For cases with $C \geq 0.25\pi^2$, SBD suppresses the amplification of $B_\phi$ throughout the entire simulation box, maintaining the physical field configuration.
Importantly, these cases yield nearly identical field distributions, indicating convergence of the solutions for sufficiently high $C$. 
For $C \leq 0.125\pi^2$, SBD is too weak to prevent the amplification of $B_\phi$, and the field configuration transitions to an unphysical geometry, similar to the case without SBD. 
Values of $C \geq 0.25\pi^2$ correspond to diffusion timescales of $\tau_{\rm diff} \lesssim 12$ orbits (equation~\eqref{eq:SBD_tau_diff}). This threshold is comparable to the amplification timescale of $B_\phi$ in the absence of SBD.
We therefore conclude that $\tau_{\rm diff}$ must be shorter than the timescale of shear-induced magnetic field amplification for SBD to stabilize the physical field-line configuration.

\subsection{Comparison with self-similar solutions} \label{ssec:Validation_Compare_with_Lesur}
We next compare results from our SBD shearing-box simulations with self-similar MHD accretion disk solutions obtained by \citet{2021A&A...650A..35L}. \citet{2021A&A...650A..35L} numerically computed self-similar structures of steady, axisymmetric disks with different values of $\beta_0$ using the parametric diffusivity model.
The self-similar solutions, available in a public repository\footnote{Lesur, G. PPDwind (https://github.com/glesur/PPDwind).}, provide vertical profiles of the density, velocity, and magnetic field for self-similar disks with various combinations of $(\beta_0, {\rm Rm}_0, {\rm Am}_0)$. 
Here, we use the results from 8 runs with different sets of $(\beta_0, {\rm Am}_0, {\rm Rm}_0)$. 
Our local simulations are carried out for the same sets of $(\beta_0, {\rm Rm}_0, {\rm Am}_0)$, using the same parametric diffusivity model.
All our runs adopt $C = \pi^2$ and attain a quasi-steady state with a stable field-line configuration. 
To enable comparison, we convert the self-similar solutions of \citet{2021A&A...650A..35L} from spherical to cylindrical coordinates following the method described in subsection 2.2 of \citet{2021A&A...650A..35L}.

\begin{figure*}
  \includegraphics[width=\textwidth]{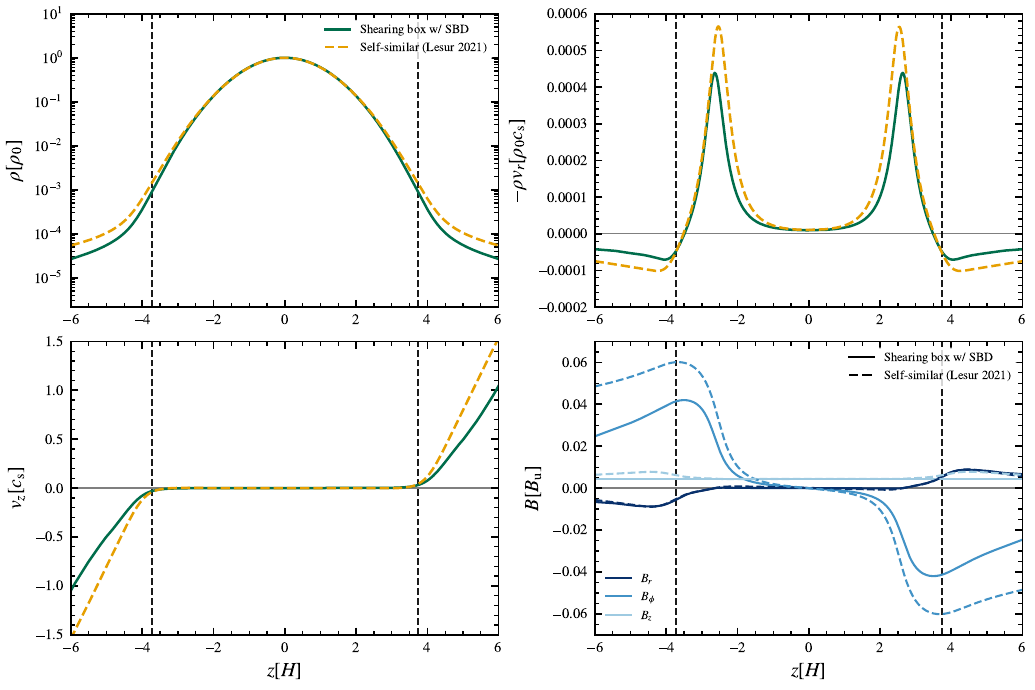}
  \caption{
  Comparison of vertical structures between shearing-box simulations (solid lines) and self-similar solutions (dashed lines) for the parametric diffusivity model with $(\beta_0, {\rm Rm}_0, {\rm Am}_0) = (10^5, 1, 1)$.
  The panels show, from upper left to lower right, the density $\rho$, the radial mass flux $-\rho v_r$, the vertical velocity $v_z$, and the magnetic field components $B_r$, $B_\phi$, and $B_z$.
  The shearing-box profiles are horizontally averaged and then time-averaged over 6 orbits in the quasi-steady state. 
  Alt text: Four line graphs arranged in two columns and two rows, sharing a common horizontal axis showing height from minus 6 to 6 scale heights. The upper left panel has a logarithmic vertical axis for density. The remaining three panels have linear vertical axes.}
\label{fig:comparison_with_lesur}
\end{figure*}

Figure~\ref{fig:comparison_with_lesur} displays the vertical disk structure for $(\beta_0, {\rm Rm}_0, {\rm Am}_0) = (10^5, 1, 1)$ from our simulation and the corresponding self-similar solution. 
The disk surface, defined as where $\beta = 1$ (see equation~\eqref{eq:def_z_surf}), lies at $|z| \equiv z_{\rm surf} \approx 3.7H$. 
In this simulation, Ohmic resistivity is sufficiently strong to decouple the magnetic field from the gas at $|z| \lesssim 2H$, resulting in accretion flows confined to narrow layers of $|z| \sim 2$--$4H$ (see also figure 13 of \citealt{2021A&A...650A..35L}). These accretion layers are associated with steep gradients in $B_\phi$, indicating that the flows are driven by the vertical gradient of the $B_\phi B_z$ Maxwell stress, related to the second term on the right-hand side of equation~\eqref{eq:mdot_stress_general}. 
At $|z| > z_{\rm surf}$, where $\beta < 1$, the magnetic field accelerates winds, with the vertical velocity increasing with height. The magnetic field and winds remove angular momentum from the accretion layers, a mechanism called MHD wind-driven accretion. 

Figure~\ref{fig:comparison_with_lesur} demonstrates that our shearing-box simulation with SBD reproduces the overall wind-driven accretion structure of the self-similar model. 
Focusing on the disk interior defined by $|z| < z_{\rm surf}$, the vertical profiles of $\rho$, $v_z$, $B_r$ and $B_z$ closely match the self-similar solution.
In contrast, the peak of the radial mass flux $-\rho v_r$ from our simulation is lower than that from the self-similar solution by 22\%. This lower accretion flux reflects the smaller $B_\phi$ above the accretion layer in our simulation.
The larger $B_\phi$ in the self-similar solution can be attributed to the radial advection term $-\partial_r(v_r B_\phi)$ included in the induction equation for $B_\phi
$ in the self-similar model\footnote{For the self-similar solution, the contribution of each radially global term in the induction equation can be calculated using the assumed radial self-similarity and the vertical disk structure. Our data analysis shows that the $-\partial_r(v_r B_\phi)$ term indeed dominates over the other global terms in the induction equation for $B_\phi$ above the accretion layer.}. The self-similar solutions of \citet{2021A&A...650A..35L} assume $v_r \propto r^{-1/2}$ and $B_\phi \propto r^{-5/4}$, giving $-\partial_r(v_r B_\phi) = (7/4) v_r B_\phi/r $. Above the accretion layer, where the wind drives $v_r > 0$,  this advection term has the same sign as $B_\phi$, thereby acting as an amplification term in the induction equation for $B_\phi$. This term is absent in our local shearing-box model, which assumes no radial variation of $v_r$ and $B_\phi$, qualitatively explaining the smaller $B_\phi$ around and above the accretion layer in our solution.

\begin{figure*}
  \includegraphics[width=\textwidth]{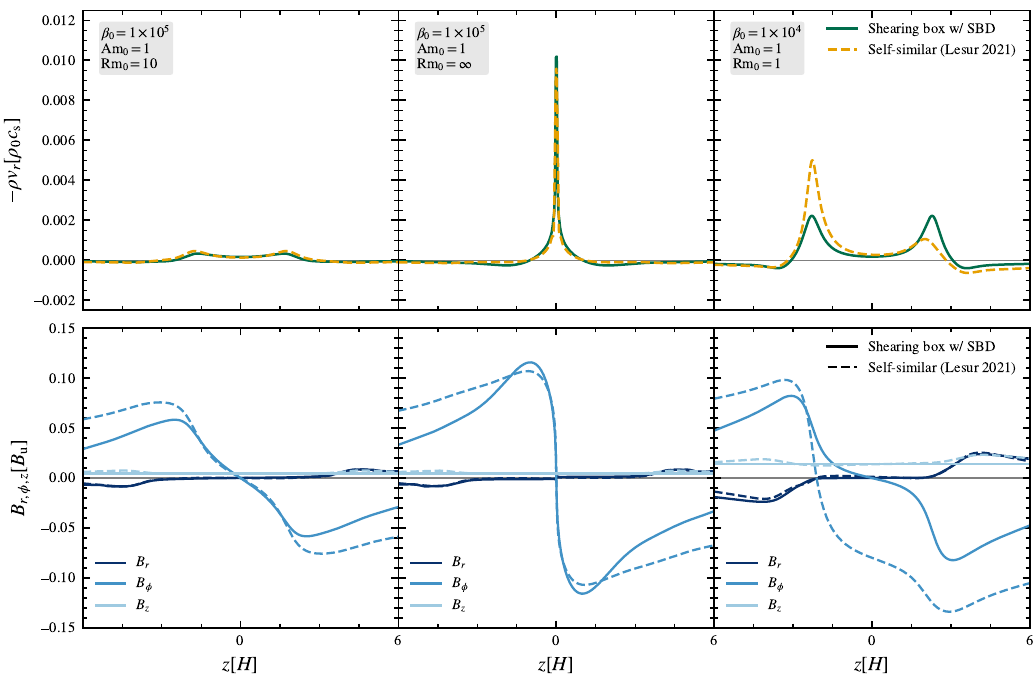}
  \caption{ 
  Comparison of the accretion and magnetic field profiles from shearing-box simulations (solid lines) and self-similar solutions (dashed lines) for different parameter sets. The left, middle, and right columns show the results for the parametric diffusivity model with $(\beta_0, {\rm Am}_0, {\rm Rm}_0) = (10^5, 1, 10)$, $(10^5, 1, \infty)$, and $(10^4, 1, 1)$, respectively.
  The upper and lower rows show the radial mass flux $-\rho v_r$ and the normalized magnetic field components $B_r$, $B_\phi$, and $B_z$, respectively. Each profile is horizontally averaged and then time-averaged over 6 orbits in the quasi-steady state.
  Alt text: Six line graphs arranged in three columns and two rows, sharing a common horizontal axis showing height from minus 6 to 6 scale heights. The upper row shows linear vertical axes for radial mass flux. The lower row shows linear vertical axes for three magnetic field components.}
\label{fig:comparison_with_lesur_other}
\end{figure*}

Figure~\ref{fig:comparison_with_lesur_other} compares the accretion and magnetic field structures from shearing-box simulations and self-similar solutions for three other cases.
As ${\rm Rm}_0$ increases, the accretion layer shifts toward the midplane, a trend consistent with the self-similar solutions. This trend occurs because weaker Ohmic dissipation (larger ${\rm Rm}_0$) allows the magnetic field to remain coupled to the gas down to deeper layers.
Even in the limit of ${\rm Rm} = \infty$, SBD maintains the physical field-line geometry. For $\beta_0 = 10^4$ (right panel), the self-similar solution is asymmetric with respect to the midplane. In this case, the peak amplitude of $-\rho v_r$ in our simulation differs from that in the self-similar solution by up to $57\%$. Nevertheless, the positions of the accretion layers predicted from the two models agree well. Moreover, the vertically integrated accretion fluxes, $\dot{M}$, obtained from these models differ by only 27\%.

\begin{figure}
  \includegraphics[width=\columnwidth]{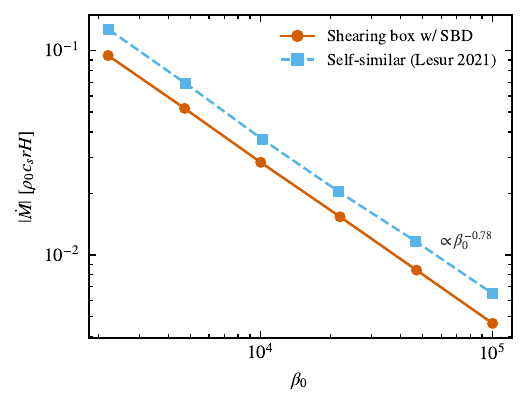}
  \caption{
  Comparison of the $\beta_0$ dependence of the mass accretion rate between shearing-box simulations (red circles, solid line) and self-similar solutions (squares, dashed line) for the parametric diffusivity model with $({\rm Rm}_0, {\rm Am}_0) = (1, 1)$. Each data point for the shearing-box simulations is obtained from a time average over 6 orbits in the quasi-steady state. Alt text: A scatter plot with logarithmic axes. The horizontal axis shows the initial plasma beta from 2.2 times 10 to the 3 to 10 to the 5. The vertical axis shows the mass accretion rate in units of rho 0 cs r H. A reference line proportional to beta 0 to the minus 0.78 is shown.}
\label{fig:beta}
\end{figure}

Figure~\ref{fig:beta} compares the mass accretion rates $\dot{M}$ given by equation~\eqref{eq:mdot_def} obtained from local simulations and self-similar solutions with $({\rm Am}_0, {\rm Rm}_0) = (1, 1)$ fixed, over the range $\beta_0 = 2.2\times10^3$ to $1.0\times10^5$. 
Both show a similar $\beta_0$ dependence. The self-similar solutions follow $\dot{M} \propto \beta_0^{-0.78}$.
The local solutions show a nearly parallel trend, although the local values are systematically smaller than the self-similar solutions by $23$--$28\%$. As discussed above, this discrepancy arises from the difference in the assumed radial disk structure for $v_r$ and $B_\phi$ between the self-similar and local shearing-box models. It is worth noting that this difference results in accretion rates that differ by only $\sim 20$--30\%.

We note that only the results for $({\rm Am}_0, {\rm Rm}_0) = (1, 1)$ are shown here; the dependence on ${\rm Am}_0$ and ${\rm Rm}_0$ is discussed in subsection~\ref{sec:results_scaling}.

%==============================================================================================================
\section{Deriving a scaling law for the accretion rate}\label{sec:Scaling}
In the previous section, we showed that local simulations with SBD reproduce the accretion rates of global self-similar solutions using a parametrized resistivity model to within 23--28\%.
In this section, we adopt a more realistic, tabulated diffusivity model presented in subsubsection~\ref{subsubsec:tabdiffmodel} to systematically investigate how the mass accretion rate varies across a wide parameter space. The goal of this section is to derive a scaling law for the accretion rate in terms of the vertical magnetic flux and resistivity distributions. 
 
\subsection{Parameter survey}\label{sec:results_survey}
We perform a total of 46 local simulations with SBD using the tabulated diffusivity model across various sets of disk parameters, including orbital radii $r = \{1, 3, 10, 30\}~{\rm au}$, surface densities $\Sigma = \{10^2, 10^3, 10^4\}~{\rm g\,cm^{-2}}$, initial plasma beta values $\beta_0 = \{3.2\times 10^3, 10^4, 3.2\times 10^4, 10^5, 3.2\times 10^5, 10^6\}$, and dust-to-gas ratios $f_{\rm dg} = \{10^{-5}, 10^{-4}, 10^{-3}\}$ (see table~\ref{tab:run_list_main} in appendix~\ref{appndix:Data}).
For each run, we analyze diagnostics computed from  vertical profiles first averaged over the horizontal directions and then time-averaged over $t = 100 \pm 3$ orbits, when our simulations have reached a quasi-steady state.
We adopt the fiducial SBD coefficient of $C = \pi^2$ for all runs.

To assess whether our modeling is biased by the choice of diffusivity model, we also run 34 simulations using the parametric diffusivity model, covering
 $\beta_0 = 10^4$--$10^6$, ${\rm Am}_0 = \{0.25, 0.5, 1, 2, 4\}$, and ${\rm Rm}_0 = \{1, 10, 100, \infty\}$.

\begin{figure}
    \includegraphics[width=\columnwidth]{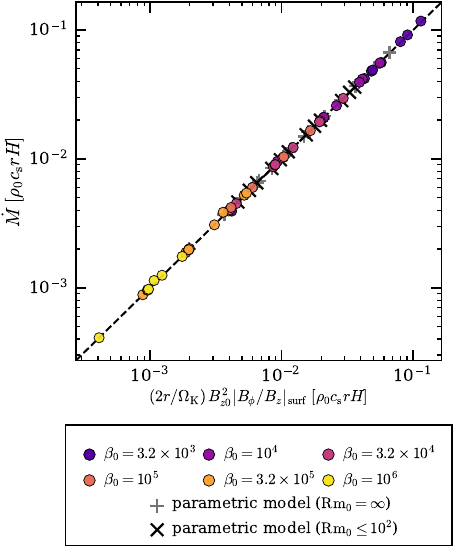}
    \caption{
    Comparison between the mass accretion rate $\dot{M}$ derived from equation~\eqref{eq:mdot_def} and the model estimate given by the right-hand side of equation~\eqref{eq:Mdot_pitch}. Both quantities are evaluated from horizontally averaged vertical profiles and then time-averaged over the quasi-steady interval. Colored circles show the 46 runs of the tabulated diffusivity model, with color indicating $\beta_0$. Black crosses and gray plus signs show the 34 runs from the parametric diffusivity model for ${\rm Rm}_0\leq10^2$ and ${\rm Rm}_0=\infty$, respectively. The dashed line shows the one-to-one relation.
    Alt text: A scatter plot with logarithmic axes. The horizontal axis shows the model estimate of the mass accretion rate in units of rho 0 cs r H. The vertical axis shows the measured mass accretion rate in units of rho 0 cs r H. Both axes span approximately t times 10 to the minus 4 to 2 times 10 to the minus 1.}
    \label{fig:mdot_vs_wind_stress}
\end{figure}

\subsection{Key diagnostics of magnetic diffusivities}\label{sec:results_effective}
Previous non-ideal MHD simulations \citep{2013ApJ...769...76B,2015ApJ...801...84G,2017A&A...600A..75B,2017ApJ...845...75B} have shown that accretion in weakly ionized protoplanetary disks is primarily driven by the $B_\phi B_z$ stress associated with MHD winds, corresponding to the second term on the right-hand side of equation~\eqref{eq:mdot_stress_general}. We therefore postulate that the accretion rates measured in our simulations can be approximated as
\begin{equation}
\dot M \approx \frac{2r}{\Omega}\,B_{z0}^2 \left|\frac{B_\phi}{B_z}\right|_{\rm surf},
\label{eq:Mdot_pitch}
\end{equation}
where $\left|B_\phi/B_z\right|_{\rm surf}$ is evaluated from the horizontally averaged vertical profiles, i.e., $\left|\langle B_\phi\rangle_{xy}/\langle B_z\rangle_{xy}\right|$ at the disk surface, after time averaging over the quasi-steady interval. We have assumed that $B_z$ is uniform in the vertical direction and approximate the value of $B_z$ at the disk surface by its midplane value $B_{z0}$ (in our shearing-box calculations, the net vertical flux is exactly vertically uniform).
Equation~\eqref{eq:Mdot_pitch} shows that the mass accretion rate scales with $B_{z0}^2$ and the field-line pitch ratio $|B_\phi/B_z|_{\rm surf}$. Figure~\ref{fig:mdot_vs_wind_stress} confirms that equation~\eqref{eq:Mdot_pitch} holds for all 46 of our tabulated model runs and 34 runs of parametric models.
This correlation indicates that the mass accretion rate can be predicted, in principle, from  $|B_\phi/B_z|_{\rm surf}$ as a function of disk parameters.

 \begin{figure}
\includegraphics[width=\columnwidth]{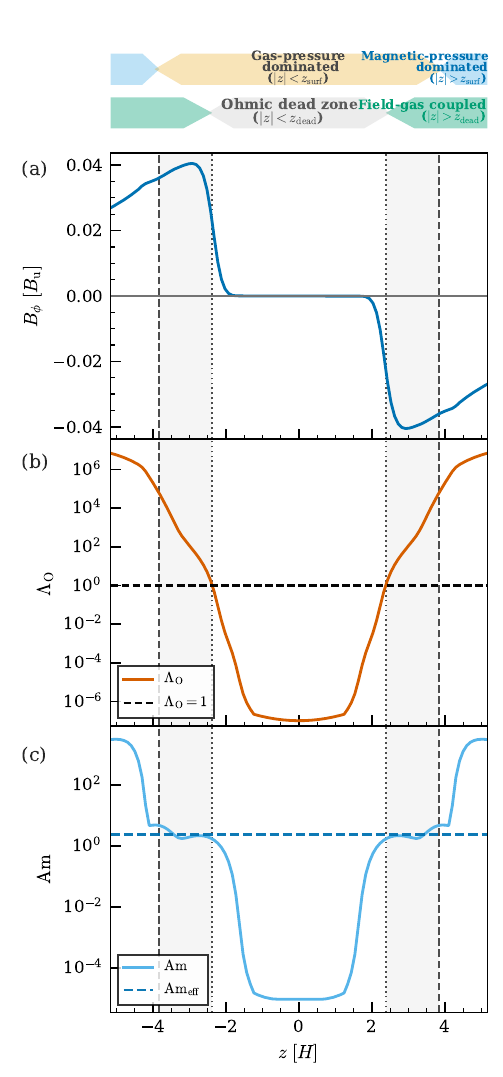}
    \caption{
    Vertical profiles of (a) the toroidal field $B_\phi$, (b) the Ohmic Elsasser number $\Lambda_{\rm O}$, and (c) the ambipolar Elsasser number ${\rm Am}$ for run \texttt{R1-B5-S3-D4}.
    The profiles are horizontally averaged and then time-averaged over the quasi-steady interval. The vertical dotted and dashed lines indicate $z_{\rm dead}$ and $z_{\rm surf}$, respectively.
    The shaded region corresponds to $z_{\rm dead} < |z| < z_{\rm surf}$.
    The horizontal dashed line in panel (b) marks $\Lambda_{\rm O} = 1$.
    The horizontal dashed line in panel (c) indicates the effective ambipolar Elsasser number ${\rm Am}_{\rm eff}$ defined in subsection~\ref{sec:results_effective}.
    Alt text: Three line graphs arranged vertically, sharing a common horizontal axis showing height from minus 5 to 5 scale heights. Panel (a) has a linear vertical axis for the toroidal field from minus 0.04 to 0.04 in units of Bu. Panels (b) and (c) have logarithmic vertical axes for the Ohmic and ambipolar Elsasser numbers, spanning 10 to the minus 7 to 10 to the 7 and 10 to the minus 5 to 10 to the x4, respectively.}
    \label{fig:Delta_important}
\end{figure}

The magnitude of $B_\phi$ should be related to the degree of gas--field coupling, as measured by the Elsasser numbers $\Lambda_{\rm O}$ and ${\rm Am}$ (equations~\eqref{eq:def_LambdaO} and~\eqref{eq:def_Am}, respectively).
To illustrate this, figure~\ref{fig:Delta_important} compares the vertical profiles of $B_\phi$, $\Lambda_{\rm O}$, and ${\rm Am}$ for run \texttt{R1-B5-S3-D4}. The figure shows that $B_\phi$ is large where $\Lambda_{\rm O} > 1$, i.e., where Ohmic resistivity is negligible.
Hereafter, we refer to the regions with $\Lambda_{\rm O} > 1$ and $\Lambda_{\rm O} < 1$ within the disk interior ($|z| < z_{\rm surf}$, or equivalently $\beta > 1$) as the magnetically active layer and the Ohmic dead zone, respectively. The boundary height $z_{\rm dead}$ between these two regions is defined by
\begin{equation}
\Lambda_{\rm O}(z_{\rm dead}) = 1,
\end{equation}
and the thickness of the active layer is given by $|z_{\rm surf} - z_{\rm dead}|$. As we show in the following subsection, the thickness of the active layer serves as a useful diagnostic for predicting $|B_\phi/B_z|_{\rm surf}$.

The thickness of the active layer, however, does not by itself specify the strength of magnetic coupling within that layer. At $|z| > z_{\rm dead}$, ambipolar diffusion generally determines the gas--field coupling. In our disk ionization model, ${\rm Am}$ tends to take nearly constant values of $\sim 1$ in the magnetically active layers, as shown in figure~\ref{fig:Delta_important} (see also figure 1 of \citealt{2013ApJ...769...76B} for a similar example). However, this tendency does not always hold in general ionization models, because the ionization fraction can vary by orders of magnitude with height owing to the attenuation of external ionizing sources such as cosmic rays and stellar UV/X-rays.
Figure 34 of \citet{2024PASJ...76..616I} provides an example in which ${\rm Am}$ exhibits steep variation in the surface region where $\Lambda_{\rm O} > 1$.
With this in mind, we define the effective ambipolar Elsasser number ${\rm Am}_{\rm eff}$ by $\log {\rm Am}_{\rm eff} = \langle \log {\rm Am} \rangle_{\rm active}$, where $\langle \log {\rm Am} \rangle_{\rm active}$ denotes the average of $\log {\rm Am}$ over the magnetically active layer.
The logarithmic average is adopted to prevent ${\rm Am}_{\rm eff}$ from being biased by values of ${\rm Am}$ at specific heights.

\subsection{Scaling the field-line pitch and accretion rate with key diagnostics}\label{sec:results_scaling}
We now derive a scaling law for the field-line pitch $\left|B_\phi/B_z\right|_{\rm surf}$ at the disk surface in terms of the midplane plasma beta $\beta_0$ and key diagnostics of magnetic diffusivities introduced in the previous subsection.

\begin{figure*}
  \includegraphics[width=\textwidth]{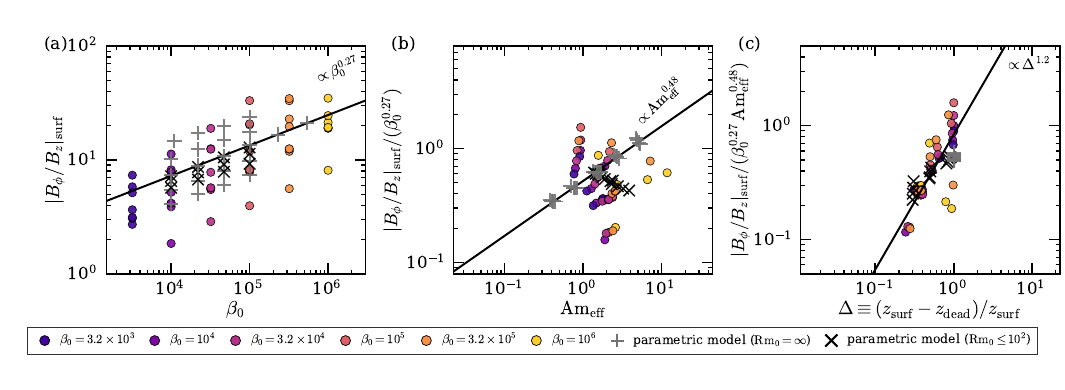}
  \caption{
    Decomposition of the parameter dependence of the field-line pitch $|B_\phi/B_z|_{\rm surf}$ on key parameters.
    Panel (a) shows the dependence on $\beta_0$, panel (b) shows the dependence on ${\rm Am}_{\rm eff}$ after dividing by $\beta_0^{0.27}$, and panel (c) shows the dependence on $\Delta$ after further dividing by ${\rm Am}_{\rm eff}^{0.48}$.
    Filled circles show the results of shearing-box simulations with SBD using the tabulated diffusivity model, with color indicating $\beta_0$.
    Plus and cross symbols show the results of shearing-box simulations with SBD using the parametric diffusivity model for ${\rm Rm}_0 = \infty$ and ${\rm Rm}_0 \le 10^2$, respectively.
    The solid lines represent the reference scalings $\propto \beta_0^{0.27}$, $\propto {\rm Am}_{\rm eff}^{0.48}$, and $\propto \Delta^{1.2}$, respectively.
    Alt text: Three scatter plots with logarithmic axes arranged horizontally. In panel (a), the horizontal axis shows beta 0 and the vertical axis shows the field-line pitch. In panel (b), the horizontal axis shows Am eff. In panel (c), the horizontal axis shows Delta.}
\label{fig:scaling_3component}
\end{figure*}

To begin with, we plot in figure~\ref{fig:scaling_3component}(a) the relation between $\left|B_\phi/B_z\right|_{\rm surf}$ and $\beta_0$ for all runs using tabulated and parametric diffusivity models.
Our results show $\left|B_\phi/B_z\right|_{\rm surf}\sim 1\text{--}20$, consistent with those from previous global simulations (e.g., \citealt{2002ApJ...581..988C,2017A&A...600A..75B,2017ApJ...845...75B}). 
Note that the scaling $\dot{M} \propto \beta_0^{-0.78}$ found by \citet{2021A&A...650A..35L} for ${\rm Rm}_0 = 1$ and ${\rm Am}_0 = 1$ (see figure~\ref{fig:beta}) is equivalent to $\left|B_\phi/B_z\right|_{\rm surf} \propto \beta_0^{0.22}$.
However, it is clear from figure~\ref{fig:scaling_3component}(a) that $\beta_0$ alone does not uniquely determine $\left|B_\phi/B_z\right|_{\rm surf}$ in more general cases. To isolate the $\beta_0$ dependence without contamination from Ohmic diffusion, we focus on runs with ${\rm Rm}_0 = \infty$ and ${\rm Am}_0 = 1$, finding $\left|B_\phi/B_z\right|_{\rm surf} \propto \beta_0^{0.27}$, which we show as the reference line in figure~\ref{fig:scaling_3component}(a).

We next plot $\left|B_\phi/B_z\right|_{\rm surf}$ normalized by $\beta_0^{0.27}$ as a function of ${\rm Am}_{\rm eff}$ (figure~\ref{fig:scaling_3component}(b)).
The plot still shows substantial scatter, indicating that ${\rm Am}_{\rm eff}$ alone does not determine $\left|B_\phi/B_z\right|_{\rm surf}/\beta_0^{0.27}$.
To isolate the effect of ambipolar diffusion, we here focus on the runs for the parametric diffusivity model with ${\rm Rm}_0 = \infty$ (plus symbols in figure~\ref{fig:scaling_3component}), where Ohmic dissipation is negligible. 
Within this restricted subset, the normalized $\left|B_\phi/B_z\right|_{\rm surf}$ increases with ${\rm Am}_{\rm eff}$, suggesting that a larger ${\rm Am}_{\rm eff}$ leads to stronger gas--field coupling in the active layer and hence stronger amplification of $B_\phi$.
A power-law fit to this subset yields $\left|B_\phi/B_z\right|_{\rm surf}/\beta_0^{0.27} \propto {\rm Am}_{\rm eff}^{0.48}$.

Finally, we isolate the effect of Ohmic resistivity by normalizing $\left|B_\phi/B_z\right|_{\rm surf}$ with $\beta_0^{0.27}{\rm Am}_{\rm eff}^{0.48}$. We characterize the effect of Ohmic resistivity using the normalized thickness of the magnetically active layer,
\begin{equation}
\Delta \equiv \frac{|z_{\rm surf}-z_{\rm dead}|}{|z_{\rm surf}|},
\label{eq:Delta_def}
\end{equation}
where the active layer vanishes when $\Delta \to 0$ and the Ohmic dead zone vanishes when $\Delta \to 1$. Figure~\ref{fig:scaling_3component}(c) plots $\left|B_\phi/B_z\right|_{\rm surf}/(\beta_0^{0.27}{\rm Am}_{\rm eff}^{0.48})$ versus $\Delta$ for all runs. 
The figure shows that the normalized $\left|B_\phi/B_z\right|_{\rm surf}$ is positively correlated with $\Delta$, suggesting that a thicker active layer leads to stronger amplification of $B_\phi$.
A power-law fit yields $\left|B_\phi/B_z\right|_{\rm surf}/(\beta_0^{0.27}{\rm Am}_{\rm eff}^{0.48}) \propto \Delta^{1.2}$.

\begin{figure*}
  \includegraphics[width=\textwidth]{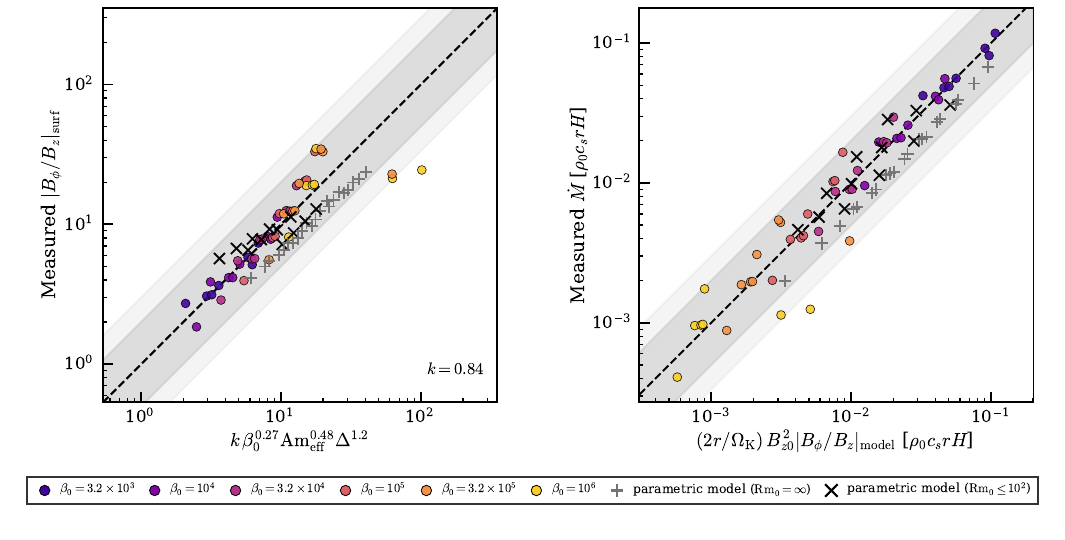}
  \caption{
    Validation of the scaling laws.
    The left panel compares the model prediction $k\beta_0^{0.27} {\rm Am}_{\rm eff}^{0.48} \Delta^{1.2}$ with the measured $|B_\phi/B_z|_{\rm surf}$, and the right panel compares the mass accretion rate derived from equation~\eqref{eq:mdot_def} with the model prediction based on equation~\eqref{eq:pitch_scaling}.
    Each data point represents the result of a shearing-box simulation with SBD.
    Filled circles show the results from the tabulated diffusivity model, and plus and cross symbols show the results from the parametric diffusivity model.
    The dashed line indicates where the model prediction and the measured value agree exactly. The dark and light gray bands indicate agreement within a factor of 2 and a factor of 3, respectively. Alt text: Two scatter plots with logarithmic axes arranged horizontally. In the left panel, the horizontal axis shows the model prediction of the field-line pitch, and the vertical axis shows the measured field-line pitch. In the right panel, the horizontal axis shows the model prediction of the mass accretion rate in units of rho, sound speed, r, and H, and the vertical axis shows the measured mass accretion rate in units of rho, sound speed, r, and H.}

\label{fig:BphiBz_Mdot}
\end{figure*}

To summarize, we have shown that 
the field-line pitch $\left|B_\phi/B_z\right|_{\rm surf}$ can be scaled as 
\begin{equation}
\left|\frac{B_\phi}{B_z}\right|_{\rm surf}
\approx k\,\beta_0^{0.27}\,{\rm Am}_{\rm eff}^{0.48}\,\Delta^{1.2},
\label{eq:pitch_scaling}
\end{equation}
where $k$ is a numerical coefficient with a best-fit value of $k = 0.84$. 
Substituting equation~\eqref{eq:pitch_scaling} into equation~\eqref{eq:Mdot_pitch}, we also obtain a scaling formula for the mass accretion rate,
\begin{equation}
\dot{M} \approx k \frac{2r}{\Omega_K}\,B_{z0}^2\beta_0^{0.27}\,{\rm Am}_{\rm eff}^{0.48}\,\Delta^{1.2}.
\label{eq:mdot_predict}
\end{equation}
Figure~\ref{fig:BphiBz_Mdot} compares the predicted and directly measured values of $\left|B_\phi/B_z\right|_{\rm surf}$ and $\dot{M}$ for all runs. We find that equations~\eqref{eq:pitch_scaling} and \eqref{eq:mdot_predict} reproduce the measured values to within a factor of three across the explored parameter space  ($\beta_0 = 3\times 10^3$--$10^6$, ${\rm Am}_{\rm eff} \sim 0.1$--$10$, $\Delta \sim 0.1$--$1$). With the exception of the least magnetized cases ($\beta_0 = 10^6$), our formulae agree with the simulation results to within a factor of 2--3. 
 
\section{Discussion}\label{sec:discussion}
\subsection{Toward global modeling of wind-driven accretion disks without $\alpha$ parameters}
The scaling formula (equation~\eqref{eq:mdot_predict}) derived in subsection~\ref{sec:results_scaling}  allows us to predict the accretion rate $\dot{M}$ from local disk quantities, namely $\beta_0$, ${\rm Am}_{\rm eff}$, and $\Delta$.
In one-dimensional disk evolution models, the magnetic diffusivity table can be evaluated locally for the evolving disk radius $r$ and surface density $\Sigma$. Equation~\eqref{eq:mdot_predict} then gives the corresponding accretion rate, allowing the scaling formula to be incorporated into global disk evolution calculations without introducing an additional $\alpha$ parameter.

Several one-dimensional disk evolution models incorporating magnetically driven accretion have been proposed in recent years \citep{2013ApJ...778L..14A,2016A&A...596A..74S,2016ApJ...821...80B,2017ApJ...845...31H,2019ApJ...879...98C,2022MNRAS.512.2290T}. 
For instance, the model of \citet{2022MNRAS.512.2290T} extends the classical viscous $\alpha$ model to include MHD wind-driven accretion and parametrizes the mass accretion rate as
\begin{equation}
    \dot{M} = 3\pi \Sigma \frac{c_s^2}{\Omega_K} \alpha_{\rm DW},
    \label{eq:mdot_alphaDW_def}
\end{equation}
where $\alpha_{\rm DW}$ is a dimensionless parameter that corresponds to the $\alpha$ parameter in the classical viscous model. 
However, since our equation~\eqref{eq:mdot_predict} gives $\dot{M}$ explicitly as a function of disk physical quantities, there is no need to parametrize $\dot{M}$ with $\alpha_{\rm DW}$ as in equation~\eqref{eq:mdot_alphaDW_def}. 
For given $B_z$, equation~\eqref{eq:mdot_predict} directly determines how the mass distribution in a wind-driven accretion disk evolves. The remaining uncertainty lies in the evolution of $B_z$, which is discussed in subsection~\ref{sec:limitations}.

Nonetheless, given that the model of \citet{2022MNRAS.512.2290T} is widely used in the literature \citep[e.g.,][]{2023ASPC..534..539M}, it is useful to analyze how the parameter $\alpha_{\rm DW}$ relates to disk properties in light of our scaling formula. Comparing  equations~\eqref{eq:Mdot_pitch} and \eqref{eq:mdot_alphaDW_def} and using $\Sigma = \sqrt{2\pi}\rho_0H$ (which follows from equation~\eqref{eq:def_rho_exp} with $\int \rho dz = \Sigma$), we can obtain
\begin{equation}
\alpha_{\text{DW}} = \frac{16}{3\sqrt{2\pi}} \frac{1}{\epsilon \beta_0} \left| \frac{B_\phi}{B_z} \right|_{\text{surf}}, \label{eq:def_alpha_DW}
\end{equation}
where $\epsilon \equiv H/r$ is the disk aspect ratio.
By substituting the empirical range of $|B_\phi/B_z|_{\rm surf}$ from global simulations into the definition of $\alpha_{\rm DW}$, \citet{2022MNRAS.512.2290T} estimate (see their equation~(69))
\begin{equation}
    \alpha_{\rm DW} \approx (0.2\text{--}4) \times 10^{-2}
    \left(\frac{\beta_0}{10^4}\right)^{-1}
    \left(\frac{\epsilon}{0.1}\right)^{-1},
    \label{eq:alpha_DW}
\end{equation}
where the range of the pre-factor reflects the uncertainty in $|B_\phi/B_z|_{\rm surf}$ \citep{2002ApJ...581..988C,2017A&A...600A..75B,2017ApJ...845...75B}.
Our scaling law for $|B_\phi/B_z|_{\rm surf}$ enables us to reduce the uncertainty in $\alpha_{\rm DW}$; substituting equation~\eqref{eq:pitch_scaling} into equation~\eqref{eq:def_alpha_DW} yields
\begin{align}
\alpha_{\rm DW} 
&\approx 
\frac{16k}{3\sqrt{2\pi}} \epsilon^{-1} \beta_0^{-0.73} {\rm Am}_{\rm eff}^{0.48} \Delta^{1.2}  
\nonumber \\
&\approx 2.1 \times 10^{-2}
\left(\frac{\beta_0}{10^4}\right)^{-0.73}
\left(\frac{\epsilon}{0.1}\right)^{-1}
{\rm Am}_{\rm eff}^{0.48} \Delta^{1.2}.
\label{eq:alpha_DW_thisstudy}
\end{align}
This expression is accurate to within a factor of 2--3 (see figure~\ref{fig:BphiBz_Mdot}).
The $\beta_0^{-0.73}$ dependence of $\alpha_{\rm DW}$ corresponds to the $B_{z0}^2\beta_0^{0.27} \propto \beta_0^{-0.73}$ dependence of $\dot{M}$ (see equation~\eqref{eq:mdot_predict}). The factors ${\rm Am}_{\rm eff}^{0.48}$ and $\Delta^{1.2}$ reflect the fact that stronger ambipolar diffusion and Ohmic diffusion each lead to a smaller $|B_\phi/B_z|_{\rm surf}$ (see section~\ref{ssec:method_accretion_rate}).

\subsection{Comparison with previous local simulations without SBD}
With SBD, local shearing-box simulations can maintain the physical field-line geometry over long periods, allowing direct measurement of $\dot{M}$.
In contrast, previous local studies without SBD indirectly estimated $\dot{M}$---by measuring the magnetic stress $B_z B_\phi$ at either the upper or lower disk surface and substituting it into equation~\eqref{eq:mdot_stress_general}---because the physical field-line geometry breaks down in some cases \citep{2013ApJ...769...76B, 2014ApJ...791..137B}.
In this section, we compare our scaling formula for $\dot{M}$ with the previous stress-based estimate by \citet{2013ApJ...772...96B,2014ApJ...791..137B}.

\citet{2014ApJ...791..137B} presented a stress-based empirical formula for $\dot{M}$ based on Hall-free shearing box simulations by \citet{2013ApJ...772...96B}, which adopts the minimum-mass solar nebula (MMSN) model of \cite{1981PThPS..70...35H}. The formula reads (see equation~(28) of \citealt{2014ApJ...791..137B})
\begin{equation}
\dot{M}= 0.47\times10^{-8}\,r_{\rm au}^{1.90}\left(\frac{B_z}{10\,{\rm mG}}\right)^{1.32}M_\odot\,{\rm yr}^{-1},
\label{eq:mdot_bai14}
\end{equation}
where $r_{\rm au} = r / 1\,{\rm au}$.
To allow direct comparison with equation~\eqref{eq:mdot_bai14}, we now rewrite our scaling law (equation~\eqref{eq:mdot_predict}), using $\rho_0 = \Sigma/(\sqrt{2\pi}H)$ together with the MMSN surface density, temperature, and rotation profiles of
$\Sigma_{\rm MMSN} = 1700\,r_{\rm au}^{-3/2}\ {\rm g\,cm^{-2}}$, $T_{\rm MMSN} = 280\,r_{\rm au}^{-1/2}\ {\rm K}$, and $\Omega_{\rm K} = 2.0\times10^{-7}\,r_{\rm au}^{-3/2}\ {\rm s^{-1}}$, as
\begin{align}
    \dot{M} &= 1.2\times10^{-8} {\rm Am}_{\rm eff}^{0.48}\,\Delta^{1.2}\,r_{\rm au}^{1.62}
    \left(\frac{B_z}{10\,{\rm mG}}\right)^{1.46}\nonumber \\
    &\times \left(\frac{\Sigma}{\Sigma_{\rm MMSN}}\right)^{0.27}
    \left(\frac{T}{T_{\rm MMSN}}\right)^{0.14}
    \,M_\odot\,{\rm yr}^{-1}.
    \label{eq:mdot_full}
\end{align}
Unlike equation~\eqref{eq:mdot_predict}, 
equation~\eqref{eq:mdot_full} accounts for the dependence on the surface density, temperature, and magnetic diffusivities.

In both formulae, the power-law index of $B_z$ is smaller than 2. 
If $|B_\phi/B_z|_{\rm surf}$ were independent of $B_z$, 
equation~\eqref{eq:Mdot_pitch} would give $\dot{M} \propto B_z^2$. 
As discussed in subsection~\ref{sec:results_scaling}, however, 
$|B_\phi/B_z|_{\rm surf} \propto \beta_0^{0.27} \propto B_z^{-0.54}$, and $\Delta$ also depends on $B_z$, both of which reduce the effective power-law index below 2. 
It is therefore natural that the explicit $B_z$ dependences in the two formulae do not match exactly.
For typical values of ${\rm Am}_{\rm eff} \sim 1$ and $\Delta \sim 0.4$--$0.5$, the prefactor $1.2\times10^{-8}{\rm Am}_{\rm eff}^{0.48}\,\Delta^{1.2}$ in our formula yields $0.4$--$0.5 \times 10^{-8}$, consistent with the prefactor $0.47 \times 10^{-8}$ in the formula of \citet{2014ApJ...791..137B}.
Our formula suggests that $\dot{M}$ depends weakly on $\Sigma$ and $T$, a dependence that arises from the $\beta_0$ dependence of $|B_\phi/B_z|_{\rm surf}$.

\subsection{Limitations of this work}
\label{sec:limitations}
The scaling law derived in this study provides a practical framework for estimating the mass accretion rate $\dot{M}$ from local disk quantities.
The largest uncertainty in applying this scaling law to global disk evolution models is the vertical magnetic field strength $B_z$, which is represented by $\beta_0$ in our simulations.
Direct observational constraints on the large-scale magnetic field threading protoplanetary disks remain limited, and current knowledge relies mainly on indirect evidence from disk observations \citep{2025ApJ...991L...6T,2025NatAs...9..526O} and from the remanent magnetization of meteorites (e.g.,~\citealt{2021SciA....7.5967W,2023Sci...379.8671N,2026JGRE..13109265S}). 
On the theoretical side, no framework has yet been established that uniquely determines the global distribution and time evolution of $B_z$. 
Global transport models based on magnetic flux advection and diffusion have long been proposed and applied to protoplanetary disks.
However, they still do not uniquely determine the distribution and time evolution of $B_z$ \citep{1994MNRAS.267..235L,2014ApJ...785..127O,2014ApJ...797..132T,2014MNRAS.441..852G}.
Furthermore, because local shearing-box simulations cannot self-consistently treat global radial magnetic flux transport, $\beta_0$ itself must be specified externally.
The scaling law derived in this study therefore characterizes the systematic dependence of $\dot{M}$ on disk conditions for a given $\beta_0$, and its application to real disks requires a global model for the distribution and evolution of $B_z$ \citep{2017ApJ...836...46B,2020ApJ...896..126G,2021A&A...650A..35L,2021MNRAS.507.1106C,2024PASJ...76..616I}.

The model adopted in this study includes several additional simplifications beyond those already noted above. We mention them here as issues to be addressed in future work.
First, our simulations include Ohmic and ambipolar diffusion but neglect the Hall effect.
The Hall effect is known to modify both the horizontal magnetic field structure and the efficiency of angular momentum transport, depending on the relative orientation between the vertical magnetic field and the disk rotation axis \citep[e.g.,][]{1999MNRAS.307..849W,2001ApJ...552..235B,2008MNRAS.385.1494K,2012MNRAS.422.2737W}.
Local simulations have shown that, when the vertical magnetic field is aligned with the disk rotation axis, the horizontal magnetic field is amplified and wind-driven accretion is enhanced, whereas in the anti-aligned case, angular momentum transport and outflow are weakened \citep{2014A&A...566A..56L,2014ApJ...791..137B}.
Recent local and global Hall-MHD simulations further demonstrate that the Hall effect can alter both the mass accretion rate, $\dot{M}$, and wind properties through modifications to the magnetic-field geometry and angular momentum transport \citep{2017A&A...600A..75B,2017ApJ...845...75B,2024MNRAS.530.5131S,2024ApJ...972..128R}.
Therefore, the accretion-rate scaling derived in this work should be regarded as a Hall-free baseline relation that may be altered in Hall-dominated regimes.

Second, as shown in equation~\eqref{eq:method_isothermal}, we assume an isothermal equation of state throughout the disk.
In reality, however, disk surface layers can be heated by irradiation and magnetic dissipation \citep{2019ApJ...872...98M,2019ApJ...874...90W,2025ApJ...992...85M}.
Such changes in the thermal structure can affect the ionization structure by modifying the vertical density distribution, and may consequently alter $z_{\rm dead}$ and $\Delta$.

Third, the tabulated diffusivity model adopted in this study is based on a specific ionization model (subsubsection~\ref{subsubsec:tabdiffmodel}).
The values of ${\rm Am}_{\rm eff}$ and $\Delta$ may therefore change if the chemical reaction network, dust grain size distribution, or treatment of ionization sources is modified. However, the accretion rate formula derived in this study, equation~\eqref{eq:mdot_predict}, is expressed in terms of the effective quantities ${\rm Am}_{\rm eff}$ and $\Delta$ determined from a given ionization structure, as long as a magnetically active layer (where $\Lambda_{\rm O}>1$ and $\beta>1$) is present above the dead zone (where $\Lambda_{\rm O}<1$) at the midplane.
We speculate that our accretion rate formula may also be applicable to other ionization models, provided that ${\rm Am}_{\rm eff}$ and $\Delta$ are re-evaluated from the corresponding ionization structure. This speculation should be tested in future work.  
We also caution that our formula is not applicable to disks that lack a magnetically active layer as defined in this study.

\section{Summary and conclusions}\label{sec:summary}
In this study, we used the super-box-scale diffusion (SBD) scheme in non-ideal MHD shearing-box simulations to quantitatively describe magnetically driven accretion in protoplanetary disks.
Local shearing-box simulations are a powerful tool for studying the vertical structure at any orbital radius with high resolution and low computational cost.
However, they suffer from a fundamental problem: the toroidal magnetic field tends to accumulate within the computational domain, making it difficult to maintain the physical field-line geometry over long periods.
In this study, we addressed this problem with SBD, thereby establishing a local framework for wind-driven accretion. 
The main results are summarized below.
\begin{enumerate}
    \item The SBD scheme (subsection~\ref{ssec:method_SBD}) suppresses the excessive accumulation of toroidal magnetic flux in the local shearing box and maintains the physical field-line geometry over long periods.
    Without SBD, $B_\phi$ accumulates in the same direction throughout the domain and the field-line geometry transitions to the unphysical configuration after approximately 110 orbits. In contrast, with SBD, the physical field-line geometry is maintained for at least 500 orbits (figures~\ref{fig:SBD_space-time_diagram} and \ref{fig:beta_75orbit_200orbit}).
    We further confirmed that, as long as $C$ is chosen large enough to keep $\tau_{\rm diff}$ shorter than the amplification timescale of $B_\phi$ by shear (figure~\ref{fig:SBD_param_C}), the quasi-steady field structure is nearly independent of the value of $C$ (figure~\ref{fig:C_Bphi_param}).    
    These results demonstrate that SBD enables a stable local framework for describing wind-driven accretion (subsection~\ref{ssec:Validation_Time_evolution}).

    \item Local simulations with SBD quantitatively agree with the self-similar solutions of \citet{2021A&A...650A..35L}.
    Comparisons using the parametric diffusivity model show that the accretion layer position and magnetic field structure agree with the self-similar solutions across a wide parameter range (figures~\ref{fig:comparison_with_lesur} and \ref{fig:comparison_with_lesur_other}).
    The $\beta_0$ dependence of the mass accretion rate also agrees with that of the self-similar solutions to within 23--28\% (figure~\ref{fig:beta}).
    This confirms that local shearing-box simulations incorporating SBD provide an effective framework for approximating the global accretion structure at low computational cost (subsection~\ref{ssec:Validation_Compare_with_Lesur}).

    \item A parameter survey of 46 runs using the magnetic diffusivity table shows that the mass accretion rate $\dot{M}$ is not uniquely determined by the net vertical magnetic field strength $\beta_0$ alone. 
    It also depends systematically on the effective parameters ${\rm Am}_{\rm eff}$ and $\Delta$ (equation~\eqref{eq:Delta_def}), which represent the strength of magnetic coupling and the normalized thickness of the magnetically active layer, respectively (figures~\ref{fig:Delta_important} and \ref{fig:scaling_3component}).
    We formulate this dependence and derive a scaling law for the field-line pitch $\left|B_\phi/B_z\right|_{\rm surf}$ (equation~(\ref{eq:pitch_scaling})) and a predictive formula for the mass accretion rate $\dot{M}$ (equation~(\ref{eq:mdot_predict})).
    These formulas reproduce the simulation results to within a factor of 2--3 across the parameter space explored in this study (figure~\ref{fig:BphiBz_Mdot}).

    \item Our scaling law determines the $\alpha_{\rm DW}$ parameter in the widely used parametric wind-driven accretion model of \citet{2022MNRAS.512.2290T} as a function of $\beta_0$, ${\rm Am}_{\rm eff}$, and $\Delta$ (equation~\eqref{eq:alpha_DW_thisstudy}).
\end{enumerate}

These results advance the description of wind-driven accretion in protoplanetary disks beyond the conventional phenomenological $\alpha$ prescription by directly linking the accretion rate to local physical quantities.
In particular, this study clarifies the physical origin of magnetically driven accretion and provides a framework for directly predicting $\dot{M}$ from $\beta_0$, ${\rm Am}_{\rm eff}$, and $\Delta$.

This study, however, has several limitations. The simulations employed in this study assume an isothermal equation of state and neglect the Hall effect. 
In addition, SBD is a prescription for incorporating large-scale magnetic relaxation into local simulations and does not self-consistently solve the radial transport of the net vertical magnetic flux.
The scaling law derived in this study therefore provides a model for predicting the local mass accretion rate for a given $B_z$. The applicability of our accretion rate formula (equation~\eqref{eq:mdot_predict}) to global disks, as well as to more general disk ionization models, should be tested in future work.

An important direction for future work is to extend the local scaling law derived here to a global evolutionary model coupled to the radial transport of the net vertical magnetic flux. 
In practice, equation~(\ref{eq:mdot_predict}) can be incorporated relatively easily into one-dimensional disk evolution calculations by updating the magnetic diffusivity table as the disk radius $r$ and surface density $\Sigma$ evolve.
The present results therefore provide a foundation for constructing long-term evolutionary models of protoplanetary disks.

\section*{Funding}
This work was supported by JST SPRING, Japan Grant Number JPMJSP2106, and by JSPS KAKENHI Grant number JP22KJ0155, JP22K14081, and JP23K25923.

\begin{ack}
We thank Xuening Bai for providing the initial numerical setup for this study, Geoffroy Lesur for making the data of his self-similar solutions publicly available, and Kazunari Iwasaki and Takeru Suzuki for discussions on global MHD accretion. We also thank the anonymous reviewer for constructive comments that helped improve the clarity of this paper. Numerical computations were carried out on HPE Cray XD2000 at the Center for Computational Astrophysics, National Astronomical Observatory of Japan.
\end{ack}

{
\appendix

\section{Data} \label{appndix:Data}
Table~\ref{tab:run_list_main} summarizes the input parameters and resulting accretion quantities for all 46 runs of the tabulated diffusivity model.

\begin{table*}[t]
\centering
\caption{\centering  List of simulations using the tabulated diffusivity model.}
\label{tab:run_list_main}
\begin{tabular}{@{}lrrrrrrrrr@{}}
\hline
Run & $r$ [au] & $\Sigma$ [g\,cm$^{-2}$] & $\log_{10}\beta_0$ & $f_{\rm dg}$ & ${\rm Am}_{\rm eff}$ & $z_{\rm dead}[H]$ & $z_{\rm surf}[H]$ & $\dot M$ [$\rho_0 c_s r H$] & $|B_\phi/B_z|_{\rm surf}$ \\
\hline
R1-B4-S2-D4 & $1$ & $10^{2}$ & $4$ & $10^{-4}$ & $1.90$ & $1.12$ & $3.19$ & $4.16\times 10^{-2}$ & $8.20$ \\
R10-B4-S3-D4 & $10$ & $10^{3}$ & $4$ & $10^{-4}$ & $0.798$ & $0.102$ & $3.21$ & $3.93\times 10^{-2}$ & $7.83$ \\
R1-B4-S3-D3 & $1$ & $10^{3}$ & $4$ & $10^{-3}$ & $1.47$ & $2.40$ & $3.57$ & $1.96\times 10^{-2}$ & $3.88$ \\
R1-B4-S3-D4 & $1$ & $10^{3}$ & $4$ & $10^{-4}$ & $1.85$ & $2.20$ & $3.56$ & $2.08\times 10^{-2}$ & $4.16$ \\
R1-B4-S3-D5 & $1$ & $10^{3}$ & $4$ & $10^{-5}$ & $1.92$ & $2.14$ & $3.56$ & $2.10\times 10^{-2}$ & $4.16$ \\
R30-B4-S3-D4 & $30$ & $10^{3}$ & $4$ & $10^{-4}$ & $0.924$ & $0.00$ & $2.96$ & $5.57\times 10^{-2}$ & $11.3$ \\
R3-B4-S3-D4 & $3$ & $10^{3}$ & $4$ & $10^{-4}$ & $1.27$ & $1.65$ & $3.45$ & $2.59\times 10^{-2}$ & $5.20$ \\
R1-B4-S4-D4 & $1$ & $10^{4}$ & $4$ & $10^{-4}$ & $1.89$ & $2.92$ & $3.85$ & $9.57\times 10^{-3}$ & $1.85$ \\
R1-B5-S2-D4 & $1$ & $10^{2}$ & $5$ & $10^{-4}$ & $2.17$ & $1.27$ & $3.32$ & $1.02\times 10^{-2}$ & $20.3$ \\
R10-B5-S3-D4 & $10$ & $10^{3}$ & $5$ & $10^{-4}$ & $0.840$ & $0.229$ & $3.31$ & $1.04\times 10^{-2}$ & $20.7$ \\
R1-B5-S3-D3 & $1$ & $10^{3}$ & $5$ & $10^{-3}$ & $2.15$ & $2.52$ & $3.83$ & $3.94\times 10^{-3}$ & $7.85$ \\
R1-B5-S3-D4 & $1$ & $10^{3}$ & $5$ & $10^{-4}$ & $2.38$ & $2.37$ & $3.83$ & $4.05\times 10^{-3}$ & $8.03$ \\
R1-B5-S3-D5 & $1$ & $10^{3}$ & $5$ & $10^{-5}$ & $2.38$ & $2.33$ & $3.84$ & $4.20\times 10^{-3}$ & $8.20$ \\
R30-B5-S3-D4 & $30$ & $10^{3}$ & $5$ & $10^{-4}$ & $0.931$ & $0.00$ & $3.03$ & $1.66\times 10^{-2}$ & $33.1$ \\
R3-B5-S3-D4 & $3$ & $10^{3}$ & $5$ & $10^{-4}$ & $1.51$ & $1.81$ & $3.62$ & $6.00\times 10^{-3}$ & $11.9$ \\
R1-B5-S4-D4 & $1$ & $10^{4}$ & $5$ & $10^{-4}$ & $2.11$ & $3.05$ & $4.17$ & $2.01\times 10^{-3}$ & $3.96$ \\
R10-B6-S3-D4 & $10$ & $10^{3}$ & $6$ & $10^{-4}$ & $6.62$ & $0.841$ & $4.01$ & $1.14\times 10^{-3}$ & $21.3$ \\
R1-B6-S3-D3 & $1$ & $10^{3}$ & $6$ & $10^{-3}$ & $2.69$ & $2.61$ & $3.97$ & $9.55\times 10^{-4}$ & $18.9$ \\
R1-B6-S3-D4 & $1$ & $10^{3}$ & $6$ & $10^{-4}$ & $2.67$ & $2.48$ & $3.97$ & $9.63\times 10^{-4}$ & $19.1$ \\
R1-B6-S3-D5 & $1$ & $10^{3}$ & $6$ & $10^{-5}$ & $2.71$ & $2.45$ & $3.97$ & $9.73\times 10^{-4}$ & $19.3$ \\
R30-B6-S3-D4 & $30$ & $10^{3}$ & $6$ & $10^{-4}$ & $11.8$ & $0.247$ & $4.08$ & $1.25\times 10^{-3}$ & $24.5$ \\
R3-B6-S3-D4 & $3$ & $10^{3}$ & $6$ & $10^{-4}$ & $1.56$ & $1.86$ & $3.65$ & $1.75\times 10^{-3}$ & $34.8$ \\
R1-B6-S4-D4 & $1$ & $10^{4}$ & $6$ & $10^{-4}$ & $2.58$ & $3.18$ & $4.38$ & $4.09\times 10^{-4}$ & $8.11$ \\
R1-B3.5-S2-D4 & $1$ & $10^{2}$ & $3.5$ & $10^{-4}$ & $1.66$ & $0.982$ & $3.05$ & $9.18\times 10^{-2}$ & $5.81$ \\
R10-B3.5-S3-D4 & $10$ & $10^{3}$ & $3.5$ & $10^{-4}$ & $0.766$ & $0.0484$ & $3.12$ & $8.13\times 10^{-2}$ & $5.15$ \\
R1-B3.5-S3-D3 & $1$ & $10^{3}$ & $3.5$ & $10^{-3}$ & $1.35$ & $2.34$ & $3.39$ & $4.21\times 10^{-2}$ & $2.72$ \\
R1-B3.5-S3-D4 & $1$ & $10^{3}$ & $3.5$ & $10^{-4}$ & $1.75$ & $2.12$ & $3.38$ & $4.79\times 10^{-2}$ & $3.07$ \\
R1-B3.5-S3-D5 & $1$ & $10^{3}$ & $3.5$ & $10^{-5}$ & $1.79$ & $2.04$ & $3.38$ & $4.90\times 10^{-2}$ & $3.14$ \\
R30-B3.5-S3-D4 & $30$ & $10^{3}$ & $3.5$ & $10^{-4}$ & $0.906$ & $0.00$ & $2.87$ & $1.18\times 10^{-1}$ & $7.34$ \\
R3-B3.5-S3-D4 & $3$ & $10^{3}$ & $3.5$ & $10^{-4}$ & $1.11$ & $1.55$ & $3.30$ & $5.60\times 10^{-2}$ & $3.65$ \\
R1-B4.5-S2-D4 & $1$ & $10^{2}$ & $4.5$ & $10^{-4}$ & $2.05$ & $1.21$ & $3.28$ & $1.97\times 10^{-2}$ & $12.5$ \\
R10-B4.5-S3-D4 & $10$ & $10^{3}$ & $4.5$ & $10^{-4}$ & $0.824$ & $0.148$ & $3.28$ & $1.94\times 10^{-2}$ & $12.4$ \\
R1-B4.5-S3-D3 & $1$ & $10^{3}$ & $4.5$ & $10^{-3}$ & $1.76$ & $2.45$ & $3.73$ & $8.69\times 10^{-3}$ & $5.49$ \\
R1-B4.5-S3-D4 & $1$ & $10^{3}$ & $4.5$ & $10^{-4}$ & $2.09$ & $2.28$ & $3.73$ & $8.98\times 10^{-3}$ & $5.66$ \\
R1-B4.5-S3-D5 & $1$ & $10^{3}$ & $4.5$ & $10^{-5}$ & $2.12$ & $2.23$ & $3.73$ & $9.02\times 10^{-3}$ & $5.69$ \\
R30-B4.5-S3-D4 & $30$ & $10^{3}$ & $4.5$ & $10^{-4}$ & $0.934$ & $0.00$ & $3.00$ & $2.95\times 10^{-2}$ & $18.9$ \\
R3-B4.5-S3-D4 & $3$ & $10^{3}$ & $4.5$ & $10^{-4}$ & $1.41$ & $1.74$ & $3.55$ & $1.23\times 10^{-2}$ & $7.79$ \\
R1-B4.5-S4-D4 & $1$ & $10^{4}$ & $4.5$ & $10^{-4}$ & $1.97$ & $2.98$ & $4.02$ & $4.50\times 10^{-3}$ & $2.87$ \\
R1-B5.5-S2-D4 & $1$ & $10^{2}$ & $5.5$ & $10^{-4}$ & $2.31$ & $1.37$ & $3.38$ & $5.22\times 10^{-3}$ & $33.1$ \\
R10-B5.5-S3-D4 & $10$ & $10^{3}$ & $5.5$ & $10^{-4}$ & $0.884$ & $0.480$ & $3.35$ & $5.43\times 10^{-3}$ & $34.5$ \\
R1-B5.5-S3-D3 & $1$ & $10^{3}$ & $5.5$ & $10^{-3}$ & $2.34$ & $2.58$ & $3.92$ & $1.87\times 10^{-3}$ & $11.9$ \\
R1-B5.5-S3-D4 & $1$ & $10^{3}$ & $5.5$ & $10^{-4}$ & $2.55$ & $2.44$ & $3.91$ & $1.96\times 10^{-3}$ & $12.4$ \\
R1-B5.5-S3-D5 & $1$ & $10^{3}$ & $5.5$ & $10^{-5}$ & $2.57$ & $2.40$ & $3.91$ & $1.97\times 10^{-3}$ & $12.5$ \\
R30-B5.5-S3-D4 & $30$ & $10^{3}$ & $5.5$ & $10^{-4}$ & $7.17$ & $0.0674$ & $3.76$ & $3.85\times 10^{-3}$ & $22.9$ \\
R3-B5.5-S3-D4 & $3$ & $10^{3}$ & $5.5$ & $10^{-4}$ & $1.59$ & $1.84$ & $3.67$ & $3.08\times 10^{-3}$ & $19.6$ \\
R1-B5.5-S4-D4 & $1$ & $10^{4}$ & $5.5$ & $10^{-4}$ & $2.40$ & $3.11$ & $4.30$ & $8.81\times 10^{-4}$ & $5.59$ \\
\hline
\end{tabular}
\end{table*}
}

\bibliographystyle{apj}
\bibliography{article}

@ARTICLE{2022A&A...658A..97D,
       author = {{Delage}, Timmy N. and {Okuzumi}, Satoshi and {Flock}, Mario and {Pinilla}, Paola and {Dzyurkevich}, Natalia},
        title = "{Steady-state accretion in magnetized protoplanetary disks}",
      journal = {\aap},
         year = 2022,
        month = feb,
       volume = {658},
          eid = {A97},
        pages = {A97},
          doi = {10.1051/0004-6361/202141689},
archivePrefix = {arXiv},
       eprint = {2110.05639},
 primaryClass = {astro-ph.EP},
       adsurl = {https://ui.adsabs.harvard.edu/abs/2022A&A...658A..97D}
}

@ARTICLE{2008ApJS..178..137S,
       author = {{Stone}, James M. and {Gardiner}, Thomas A. and {Teuben}, Peter and {Hawley}, John F. and {Simon}, Jacob B.},
        title = "{Athena: A New Code for Astrophysical MHD}",
      journal = {\apjs},
         year = 2008,
        month = sep,
       volume = {178},
       number = {1},
        pages = {137-177},
          doi = {10.1086/588755},
archivePrefix = {arXiv},
       eprint = {0804.0402},
 primaryClass = {astro-ph},
       adsurl = {https://ui.adsabs.harvard.edu/abs/2008ApJS..178..137S}
}

@ARTICLE{2012MNRAS.422.1140G,
       author = {{Gressel}, Oliver and {Nelson}, Richard P. and {Turner}, Neal J.},
        title = "{Dead zones as safe havens for planetesimals: influence of disc mass and external magnetic field}",
      journal = {\mnras},
         year = 2012,
        month = may,
       volume = {422},
       number = {2},
        pages = {1140-1159},
          doi = {10.1111/j.1365-2966.2012.20701.x},
archivePrefix = {arXiv},
       eprint = {1202.0771},
 primaryClass = {astro-ph.EP},
       adsurl = {https://ui.adsabs.harvard.edu/abs/2012MNRAS.422.1140G}
}

@ARTICLE{2013ApJ...769...76B,
       author = {{Bai}, Xue-Ning and {Stone}, James M.},
        title = "{Wind-driven Accretion in Protoplanetary Disks. I. Suppression of the Magnetorotational Instability and Launching of the Magnetocentrifugal Wind}",
      journal = {\apj},
         year = 2013,
        month = may,
       volume = {769},
       number = {1},
          eid = {76},
        pages = {76},
          doi = {10.1088/0004-637X/769/1/76},
archivePrefix = {arXiv},
       eprint = {1301.0318},
 primaryClass = {astro-ph.EP},
       adsurl = {https://ui.adsabs.harvard.edu/abs/2013ApJ...769...76B}
}

@ARTICLE{2017ApJ...845...75B,
       author = {{Bai}, Xue-Ning},
        title = "{Global Simulations of the Inner Regions of Protoplanetary Disks with Comprehensive Disk Microphysics}",
      journal = {\apj},
         year = 2017,
        month = aug,
       volume = {845},
       number = {1},
          eid = {75},
        pages = {75},
          doi = {10.3847/1538-4357/aa7dda},
archivePrefix = {arXiv},
       eprint = {1707.00729},
 primaryClass = {astro-ph.EP},
       adsurl = {https://ui.adsabs.harvard.edu/abs/2017ApJ...845...75B}
}

@ARTICLE{2017A&A...600A..75B,
       author = {{B{\'e}thune}, William and {Lesur}, Geoffroy and {Ferreira}, Jonathan},
        title = "{Global simulations of protoplanetary disks with net magnetic flux. I. Non-ideal MHD case}",
      journal = {\aap},
         year = 2017,
        month = apr,
       volume = {600},
          eid = {A75},
        pages = {A75},
          doi = {10.1051/0004-6361/201630056},
archivePrefix = {arXiv},
       eprint = {1612.00883},
 primaryClass = {astro-ph.EP},
       adsurl = {https://ui.adsabs.harvard.edu/abs/2017A&A...600A..75B}
}

@ARTICLE{2015ApJ...801...84G,
       author = {{Gressel}, Oliver and {Turner}, Neal J. and {Nelson}, Richard P. and {McNally}, Colin P.},
        title = "{Global Simulations of Protoplanetary Disks With Ohmic Resistivity and Ambipolar Diffusion}",
      journal = {\apj},
         year = 2015,
        month = mar,
       volume = {801},
       number = {2},
          eid = {84},
        pages = {84},
          doi = {10.1088/0004-637X/801/2/84},
archivePrefix = {arXiv},
       eprint = {1501.05431},
 primaryClass = {astro-ph.EP},
       adsurl = {https://ui.adsabs.harvard.edu/abs/2015ApJ...801...84G}
}

@ARTICLE{2019MNRAS.484..107S,
       author = {{Suriano}, Scott S. and {Li}, Zhi-Yun and {Krasnopolsky}, Ruben and {Suzuki}, Takeru K. and {Shang}, Hsien},
        title = "{The formation of rings and gaps in wind-launching non-ideal MHD discs: three-dimensional simulations}",
      journal = {\mnras},
         year = 2019,
        month = mar,
       volume = {484},
       number = {1},
        pages = {107-124},
          doi = {10.1093/mnras/sty3502},
archivePrefix = {arXiv},
       eprint = {1810.02234},
 primaryClass = {astro-ph.SR},
       adsurl = {https://ui.adsabs.harvard.edu/abs/2019MNRAS.484..107S}
}

@ARTICLE{2008ApJ...679L.131T,
       author = {{Turner}, N.~J. and {Sano}, T.},
        title = "{Dead Zone Accretion Flows in Protostellar Disks}",
      journal = {\apjl},
         year = 2008,
        month = jun,
       volume = {679},
       number = {2},
        pages = {L131},
          doi = {10.1086/589540},
archivePrefix = {arXiv},
       eprint = {0804.2916},
 primaryClass = {astro-ph},
       adsurl = {https://ui.adsabs.harvard.edu/abs/2008ApJ...679L.131T}
}

@ARTICLE{2021A&A...650A..35L,
       author = {{Lesur}, Geoffroy R.~J.},
        title = "{Systematic description of wind-driven protoplanetary discs}",
      journal = {\aap},
         year = 2021,
        month = jun,
       volume = {650},
          eid = {A35},
        pages = {A35},
          doi = {10.1051/0004-6361/202040109},
archivePrefix = {arXiv},
       eprint = {2101.10349},
 primaryClass = {astro-ph.SR},
       adsurl = {https://ui.adsabs.harvard.edu/abs/2021A&A...650A..35L}
}

@ARTICLE{2018MNRAS.477.1239S,
       author = {{Suriano}, Scott S. and {Li}, Zhi-Yun and {Krasnopolsky}, Ruben and {Shang}, Hsien},
        title = "{The formation of rings and gaps in magnetically coupled disc-wind systems: ambipolar diffusion and reconnection}",
      journal = {\mnras},
         year = 2018,
        month = jun,
       volume = {477},
       number = {1},
        pages = {1239-1257},
          doi = {10.1093/mnras/sty717},
archivePrefix = {arXiv},
       eprint = {1712.06217},
 primaryClass = {astro-ph.SR},
       adsurl = {https://ui.adsabs.harvard.edu/abs/2018MNRAS.477.1239S}
}

@ARTICLE{1974MNRAS.168..603L,
       author = {{Lynden-Bell}, D. and {Pringle}, J.~E.},
        title = "{The evolution of viscous discs and the origin of the nebular variables.}",
      journal = {\mnras},
         year = 1974,
        month = sep,
       volume = {168},
        pages = {603-637},
          doi = {10.1093/mnras/168.3.603},
       adsurl = {https://ui.adsabs.harvard.edu/abs/1974MNRAS.168..603L}
}

@INPROCEEDINGS{2023ASPC..534..539M,
       author = {{Manara}, C.~F. and {Ansdell}, M. and {Rosotti}, G.~P. and {Hughes}, A.~M. and {Armitage}, P.~J. and {Lodato}, G. and {Williams}, J.~P.},
        title = "{Demographics of Young Stars and their Protoplanetary Disks: Lessons Learned on Disk Evolution and its Connection to Planet Formation}",
    booktitle = {Protostars and Planets VII},
         year = 2023,
       editor = {{Inutsuka}, S. and {Aikawa}, Y. and {Muto}, T. and {Tomida}, K. and {Tamura}, M.},
       series = {Astronomical Society of the Pacific Conference Series},
       volume = {534},
        month = jul,
        pages = {539},
          doi = {10.48550/arXiv.2203.09930},
archivePrefix = {arXiv},
       eprint = {2203.09930},
 primaryClass = {astro-ph.SR},
       adsurl = {https://ui.adsabs.harvard.edu/abs/2023ASPC..534..539M}
}

@ARTICLE{1973A&A....24..337S,
       author = {{Shakura}, N.~I. and {Sunyaev}, R.~A.},
        title = "{Black holes in binary systems. Observational appearance.}",
      journal = {\aap},
         year = 1973,
        month = jan,
       volume = {24},
        pages = {337-355},
       adsurl = {https://ui.adsabs.harvard.edu/abs/1973A&A....24..337S}
}

@ARTICLE{1982MNRAS.199..883B,
       author = {{Blandford}, R.~D. and {Payne}, D.~G.},
        title = "{Hydromagnetic flows from accretion disks and the production of radio jets.}",
      journal = {\mnras},
         year = 1982,
        month = jun,
       volume = {199},
        pages = {883-903},
          doi = {10.1093/mnras/199.4.883},
       adsurl = {https://ui.adsabs.harvard.edu/abs/1982MNRAS.199..883B}
}

@ARTICLE{1989ApJ...342..208K,
       author = {{Konigl}, Arieh},
        title = "{Self-similar Models of Magnetized Accretion Disks}",
      journal = {\apj},
         year = 1989,
        month = jul,
       volume = {342},
        pages = {208},
          doi = {10.1086/167585},
       adsurl = {https://ui.adsabs.harvard.edu/abs/1989ApJ...342..208K}
}

@ARTICLE{1991ApJ...376..214B,
       author = {{Balbus}, Steven A. and {Hawley}, John F.},
        title = "{A Powerful Local Shear Instability in Weakly Magnetized Disks. I. Linear Analysis}",
      journal = {\apj},
         year = 1991,
        month = jul,
       volume = {376},
        pages = {214},
          doi = {10.1086/170270},
       adsurl = {https://ui.adsabs.harvard.edu/abs/1991ApJ...376..214B}
}

@ARTICLE{2000ApJ...543..486S,
       author = {{Sano}, Takayoshi and {Miyama}, Shoken M. and {Umebayashi}, Toyoharu and {Nakano}, Takenori},
        title = "{Magnetorotational Instability in Protoplanetary Disks. II. Ionization State and Unstable Regions}",
      journal = {\apj},
         year = 2000,
        month = nov,
       volume = {543},
       number = {1},
        pages = {486-501},
          doi = {10.1086/317075},
archivePrefix = {arXiv},
       eprint = {astro-ph/0005464},
 primaryClass = {astro-ph},
       adsurl = {https://ui.adsabs.harvard.edu/abs/2000ApJ...543..486S}
}

@ARTICLE{2007Ap&SS.311...35W,
       author = {{Wardle}, Mark},
        title = "{Magnetic fields in protoplanetary disks}",
      journal = {\apss},
         year = 2007,
        month = oct,
       volume = {311},
       number = {1-3},
        pages = {35-45},
          doi = {10.1007/s10509-007-9575-8},
archivePrefix = {arXiv},
       eprint = {0704.0970},
 primaryClass = {astro-ph},
       adsurl = {https://ui.adsabs.harvard.edu/abs/2007Ap&SS.311...35W}
}

@INPROCEEDINGS{2014prpl.conf..411T,
       author = {{Turner}, N.~J. and {Fromang}, S. and {Gammie}, C. and {Klahr}, H. and {Lesur}, G. and {Wardle}, M. and {Bai}, X.-N.},
        title = "{Transport and Accretion in Planet-Forming Disks}",
    booktitle = {Protostars and Planets VI},
         year = 2014,
       editor = {{Beuther}, Henrik and {Klessen}, Ralf S. and {Dullemond}, Cornelis P. and {Henning}, Thomas},
        month = jan,
        pages = {411-432},
          doi = {10.2458/azu_uapress_9780816531240-ch018},
archivePrefix = {arXiv},
       eprint = {1401.7306},
 primaryClass = {astro-ph.EP},
       adsurl = {https://ui.adsabs.harvard.edu/abs/2014prpl.conf..411T}
}

@ARTICLE{2014A&A...566A..56L,
       author = {{Lesur}, Geoffroy and {Kunz}, Matthew W. and {Fromang}, S{\'e}bastien},
        title = "{Thanatology in protoplanetary discs. The combined influence of Ohmic, Hall, and ambipolar diffusion on dead zones}",
      journal = {\aap},
         year = 2014,
        month = jun,
       volume = {566},
          eid = {A56},
        pages = {A56},
          doi = {10.1051/0004-6361/201423660},
archivePrefix = {arXiv},
       eprint = {1402.4133},
 primaryClass = {astro-ph.SR},
       adsurl = {https://ui.adsabs.harvard.edu/abs/2014A&A...566A..56L}
}

@ARTICLE{2013ApJ...775...73S,
       author = {{Simon}, Jacob B. and {Bai}, Xue-Ning and {Armitage}, Philip J. and {Stone}, James M. and {Beckwith}, Kris},
        title = "{Turbulence in the Outer Regions of Protoplanetary Disks. II. Strong Accretion Driven by a Vertical Magnetic Field}",
      journal = {\apj},
         year = 2013,
        month = sep,
       volume = {775},
       number = {1},
          eid = {73},
        pages = {73},
          doi = {10.1088/0004-637X/775/1/73},
archivePrefix = {arXiv},
       eprint = {1306.3222},
 primaryClass = {astro-ph.SR},
       adsurl = {https://ui.adsabs.harvard.edu/abs/2013ApJ...775...73S}
}

@ARTICLE{2020ApJ...896..126G,
       author = {{Gressel}, Oliver and {Ramsey}, Jon P. and {Brinch}, Christian and {Nelson}, Richard P. and {Turner}, Neal J. and {Bruderer}, Simon},
        title = "{Global Hydromagnetic Simulations of Protoplanetary Disks with Stellar Irradiation and Simplified Thermochemistry}",
      journal = {\apj},
         year = 2020,
        month = jun,
       volume = {896},
       number = {2},
          eid = {126},
        pages = {126},
          doi = {10.3847/1538-4357/ab91b7},
archivePrefix = {arXiv},
       eprint = {2005.03431},
 primaryClass = {astro-ph.EP},
       adsurl = {https://ui.adsabs.harvard.edu/abs/2020ApJ...896..126G}
}

@ARTICLE{2022MNRAS.512.2290T,
       author = {{Tabone}, Beno{\^\i}t and {Rosotti}, Giovanni P. and {Cridland}, Alexander J. and {Armitage}, Philip J. and {Lodato}, Giuseppe},
        title = "{Secular evolution of MHD wind-driven discs: analytical solutions in the expanded {\ensuremath{\alpha}}-framework}",
      journal = {\mnras},
         year = 2022,
        month = may,
       volume = {512},
       number = {2},
        pages = {2290-2309},
          doi = {10.1093/mnras/stab3442},
archivePrefix = {arXiv},
       eprint = {2111.10145},
 primaryClass = {astro-ph.SR},
       adsurl = {https://ui.adsabs.harvard.edu/abs/2022MNRAS.512.2290T}
}

@ARTICLE{2023Sci...379.8671N,
       author = {{Nakamura}, T. and {Matsumoto}, M. and {Amano}, K. and {Enokido}, Y. and {Zolensky}, M.~E. and {Mikouchi}, T. and {Genda}, H. and {Tanaka}, S. and {Zolotov}, M.~Y. and {Kurosawa}, K. and {Wakita}, S. and {Hyodo}, R. and {Nagano}, H. and {Nakashima}, D. and {Takahashi}, Y. and {Fujioka}, Y. and {Kikuiri}, M. and {Kagawa}, E. and {Matsuoka}, M. and {Brearley}, A.~J. and {Tsuchiyama}, A. and {Uesugi}, M. and {Matsuno}, J. and {Kimura}, Y. and {Sato}, M. and {Milliken}, R.~E. and {Tatsumi}, E. and {Sugita}, S. and {Hiroi}, T. and {Kitazato}, K. and {Brownlee}, D. and {Joswiak}, D.~J. and {Takahashi}, M. and {Ninomiya}, K. and {Takahashi}, T. and {Osawa}, T. and {Terada}, K. and {Brenker}, F.~E. and {Tkalcec}, B.~J. and {Vincze}, L. and {Brunetto}, R. and {Al{\'e}on-Toppani}, A. and {Chan}, Q.~H.~S. and {Roskosz}, M. and {Viennet}, J.-C. and {Beck}, P. and {Alp}, E.~E. and {Michikami}, T. and {Nagaashi}, Y. and {Tsuji}, T. and {Ino}, Y. and {Martinez}, J. and {Han}, J. and {Dolocan}, A. and {Bodnar}, R.~J. and {Tanaka}, M. and {Yoshida}, H. and {Sugiyama}, K. and {King}, A.~J. and {Fukushi}, K. and {Suga}, H. and {Yamashita}, S. and {Kawai}, T. and {Inoue}, K. and {Nakato}, A. and {Noguchi}, T. and {Vilas}, F. and {Hendrix}, A.~R. and {Jaramillo-Correa}, C. and {Domingue}, D.~L. and {Dominguez}, G. and {Gainsforth}, Z. and {Engrand}, C. and {Duprat}, J. and {Russell}, S.~S. and {Bonato}, E. and {Ma}, C. and {Kawamoto}, T. and {Wada}, T. and {Watanabe}, S. and {Endo}, R. and {Enju}, S. and {Riu}, L. and {Rubino}, S. and {Tack}, P. and {Takeshita}, S. and {Takeichi}, Y. and {Takeuchi}, A. and {Takigawa}, A. and {Takir}, D. and {Tanigaki}, T. and {Taniguchi}, A. and {Tsukamoto}, K. and {Yagi}, T. and {Yamada}, S. and {Yamamoto}, K. and {Yamashita}, Y. and {Yasutake}, M. and {Uesugi}, K. and {Umegaki}, I. and {Chiu}, I. and {Ishizaki}, T. and {Okumura}, S. and {Palomba}, E. and {Pilorget}, C. and {Potin}, S.~M. and {Alasli}, A. and {Anada}, S. and {Araki}, Y. and {Sakatani}, N. and {Schultz}, C. and {Sekizawa}, O. and {Sitzman}, S.~D. and {Sugiura}, K. and {Sun}, M. and {Dartois}, E. and {De Pauw}, E. and {Dionnet}, Z. and {Djouadi}, Z. and {Falkenberg}, G. and {Fujita}, R. and {Fukuma}, T. and {Gearba}, I.~R. and {Hagiya}, K. and {Hu}, M.~Y. and {Kato}, T. and {Kawamura}, T. and {Kimura}, M. and {Kubo}, M.~K. and {Langenhorst}, F. and {Lantz}, C. and {Lavina}, B. and {Lindner}, M. and {Zhao}, J. and {Vekemans}, B. and {Baklouti}, D. and {Bazi}, B. and {Borondics}, F. and {Nagasawa}, S. and {Nishiyama}, G. and {Nitta}, K. and {Mathurin}, J. and {Matsumoto}, T. and {Mitsukawa}, I. and {Miura}, H. and {Miyake}, A. and {Miyake}, Y. and {Yurimoto}, H. and {Okazaki}, R. and {Yabuta}, H. and {Naraoka}, H. and {Sakamoto}, K. and {Tachibana}, S. and {Connolly}, H.~C. and {Lauretta}, D.~S. and {Yoshitake}, M. and {Yoshikawa}, M. and {Yoshikawa}, K. and {Yoshihara}, K. and {Yokota}, Y. and {Yogata}, K. and {Yano}, H. and {Yamamoto}, Y. and {Yamamoto}, D. and {Yamada}, M. and {Yamada}, T. and {Yada}, T. and {Wada}, K. and {Usui}, T. and {Tsukizaki}, R. and {Terui}, F. and {Takeuchi}, H. and {Takei}, Y. and {Iwamae}, A. and {Soejima}, H. and {Shirai}, K. and {Shimaki}, Y. and {Senshu}, H. and {Sawada}, H. and {Saiki}, T. and {Ozaki}, M. and {Ono}, G. and {Okada}, T. and {Ogawa}, N. and {Ogawa}, K. and {Noguchi}, R. and {Noda}, H. and {Nishimura}, M. and {Namiki}, N. and {Nakazawa}, S. and {Morota}, T. and {Miyazaki}, A. and {Miura}, A. and {Mimasu}, Y. and {Matsumoto}, K. and {Kumagai}, K. and {Kouyama}, T. and {Kikuchi}, S. and {Kawahara}, K. and {Kameda}, S.},
        title = "{Formation and evolution of carbonaceous asteroid Ryugu: Direct evidence from returned samples}",
      journal = {Science},
         year = 2023,
        month = mar,
       volume = {379},
       number = {6634},
          eid = {abn8671},
        pages = {abn8671},
          doi = {10.1126/science.abn8671},
       adsurl = {https://ui.adsabs.harvard.edu/abs/2023Sci...379.8671N}
}

@ARTICLE{2021SciA....7.5967W,
       author = {{Weiss}, Benjamin P. and {Bai}, Xue-Ning and {Fu}, Roger R.},
        title = "{History of the solar nebula from meteorite paleomagnetism}",
      journal = {Science Advances},
         year = 2021,
        month = jan,
       volume = {7},
       number = {1},
        pages = {eaba5967},
          doi = {10.1126/sciadv.aba5967},
archivePrefix = {arXiv},
       eprint = {2103.02011},
 primaryClass = {astro-ph.EP},
       adsurl = {https://ui.adsabs.harvard.edu/abs/2021SciA....7.5967W}
}

@ARTICLE{1994MNRAS.267..235L,
       author = {{Lubow}, S.~H. and {Papaloizou}, J.~C.~B. and {Pringle}, J.~E.},
        title = "{Magnetic field dragging in accretion discs}",
      journal = {\mnras},
         year = 1994,
        month = mar,
       volume = {267},
       number = {2},
        pages = {235-240},
          doi = {10.1093/mnras/267.2.235},
       adsurl = {https://ui.adsabs.harvard.edu/abs/1994MNRAS.267..235L}
}

@ARTICLE{2014ApJ...785..127O,
       author = {{Okuzumi}, Satoshi and {Takeuchi}, Taku and {Muto}, Takayuki},
        title = "{Radial Transport of Large-scale Magnetic Fields in Accretion Disks. I. Steady Solutions and an Upper Limit on the Vertical Field Strength}",
      journal = {\apj},
         year = 2014,
        month = apr,
       volume = {785},
       number = {2},
          eid = {127},
        pages = {127},
          doi = {10.1088/0004-637X/785/2/127},
archivePrefix = {arXiv},
       eprint = {1310.7446},
 primaryClass = {astro-ph.EP},
       adsurl = {https://ui.adsabs.harvard.edu/abs/2014ApJ...785..127O}
}

@ARTICLE{2014ApJ...797..132T,
       author = {{Takeuchi}, Taku and {Okuzumi}, Satoshi},
        title = "{Radial Transport of Large-scale Magnetic Fields in Accretion Disks. II. Relaxation to Steady States}",
      journal = {\apj},
         year = 2014,
        month = dec,
       volume = {797},
       number = {2},
          eid = {132},
        pages = {132},
          doi = {10.1088/0004-637X/797/2/132},
archivePrefix = {arXiv},
       eprint = {1310.7380},
 primaryClass = {astro-ph.EP},
       adsurl = {https://ui.adsabs.harvard.edu/abs/2014ApJ...797..132T}
}

@ARTICLE{2014MNRAS.441..852G,
       author = {{Guilet}, J{\'e}r{\^o}me and {Ogilvie}, Gordon I.},
        title = "{Global evolution of the magnetic field in a thin disc and its consequences for protoplanetary systems}",
      journal = {\mnras},
         year = 2014,
        month = jun,
       volume = {441},
       number = {1},
        pages = {852-868},
          doi = {10.1093/mnras/stu532},
archivePrefix = {arXiv},
       eprint = {1403.3732},
 primaryClass = {astro-ph.EP},
       adsurl = {https://ui.adsabs.harvard.edu/abs/2014MNRAS.441..852G}
}

@ARTICLE{2017ApJ...836...46B,
       author = {{Bai}, Xue-Ning and {Stone}, James M.},
        title = "{Hall Effect-Mediated Magnetic Flux Transport in Protoplanetary Disks}",
      journal = {\apj},
         year = 2017,
        month = feb,
       volume = {836},
       number = {1},
          eid = {46},
        pages = {46},
          doi = {10.3847/1538-4357/836/1/46},
archivePrefix = {arXiv},
       eprint = {1612.03912},
 primaryClass = {astro-ph.EP},
       adsurl = {https://ui.adsabs.harvard.edu/abs/2017ApJ...836...46B}
}

@ARTICLE{2025NatAs...9..526O,
       author = {{Ohashi}, Satoshi and {Muto}, Takayuki and {Tsukamoto}, Yusuke and {Kataoka}, Akimasa and {Tsukagoshi}, Takashi and {Momose}, Munetake and {Fukagawa}, Misato and {Sakai}, Nami},
        title = "{Observationally derived magnetic field strength and 3D components in the HD 142527 disk}",
      journal = {Nature Astronomy},
         year = 2025,
        month = apr,
       volume = {9},
        pages = {526-534},
          doi = {10.1038/s41550-024-02454-x},
archivePrefix = {arXiv},
       eprint = {2502.06030},
 primaryClass = {astro-ph.EP},
       adsurl = {https://ui.adsabs.harvard.edu/abs/2025NatAs...9..526O}
}

@ARTICLE{2019ApJ...872...98M,
       author = {{Mori}, Shoji and {Bai}, Xue-Ning and {Okuzumi}, Satoshi},
        title = "{Temperature Structure in the Inner Regions of Protoplanetary Disks: Inefficient Accretion Heating Controlled by Nonideal Magnetohydrodynamics}",
      journal = {\apj},
         year = 2019,
        month = feb,
       volume = {872},
       number = {1},
          eid = {98},
        pages = {98},
          doi = {10.3847/1538-4357/ab0022},
archivePrefix = {arXiv},
       eprint = {1901.06921},
 primaryClass = {astro-ph.EP},
       adsurl = {https://ui.adsabs.harvard.edu/abs/2019ApJ...872...98M}
}

@ARTICLE{2019ApJ...874...90W,
       author = {{Wang}, Lile and {Bai}, Xue-Ning and {Goodman}, Jeremy},
        title = "{Global Simulations of Protoplanetary Disk Outflows with Coupled Non-ideal Magnetohydrodynamics and Consistent Thermochemistry}",
      journal = {\apj},
         year = 2019,
        month = mar,
       volume = {874},
       number = {1},
          eid = {90},
        pages = {90},
          doi = {10.3847/1538-4357/ab06fd},
archivePrefix = {arXiv},
       eprint = {1810.12330},
 primaryClass = {astro-ph.EP},
       adsurl = {https://ui.adsabs.harvard.edu/abs/2019ApJ...874...90W}
}

@INPROCEEDINGS{2023ASPC..534..465L,
       author = {{Lesur}, G. and {Flock}, M. and {Ercolano}, B. and {Lin}, M.-K. and {Yang}, C. and {Barranco}, J.~A. and {Benitez-Llambay}, P. and {Goodman}, J. and {Johansen}, A. and {Klahr}, H. and {Laibe}, G. and {Lyra}, W. and {Marcus}, P.~S. and {Nelson}, R.~P. and {Squire}, J. and {Simon}, J.~B. and {Turner}, N.~J. and {Umurhan}, O.~M. and {Youdin}, A.~N.},
        title = "{Hydro-, Magnetohydro-, and Dust-Gas Dynamics of Protoplanetary Disks}",
    booktitle = {Protostars and Planets VII},
         year = 2023,
       editor = {{Inutsuka}, S. and {Aikawa}, Y. and {Muto}, T. and {Tomida}, K. and {Tamura}, M.},
       series = {Astronomical Society of the Pacific Conference Series},
       volume = {534},
        month = jul,
        pages = {465},
          doi = {10.48550/arXiv.2203.09821},
archivePrefix = {arXiv},
       eprint = {2203.09821},
 primaryClass = {astro-ph.EP},
       adsurl = {https://ui.adsabs.harvard.edu/abs/2023ASPC..534..465L}
}

@ARTICLE{2021MNRAS.507.1106C,
       author = {{Cui}, Can and {Bai}, Xue-Ning},
        title = "{Global three-dimensional simulations of outer protoplanetary discs with ambipolar diffusion}",
      journal = {\mnras},
         year = 2021,
        month = oct,
       volume = {507},
       number = {1},
        pages = {1106-1126},
          doi = {10.1093/mnras/stab2220},
archivePrefix = {arXiv},
       eprint = {2106.10167},
 primaryClass = {astro-ph.EP},
       adsurl = {https://ui.adsabs.harvard.edu/abs/2021MNRAS.507.1106C}
}

@ARTICLE{2019PASJ...71..100S,
       author = {{Suzuki}, Takeru K. and {Taki}, Tetsuo and {Suriano}, Scott S.},
        title = "{Magnetohydrodynamics in a cylindrical shearing box}",
      journal = {\pasj},
         year = 2019,
        month = oct,
       volume = {71},
       number = {5},
          eid = {100},
        pages = {100},
          doi = {10.1093/pasj/psz082},
archivePrefix = {arXiv},
       eprint = {1904.05032},
 primaryClass = {astro-ph.HE},
       adsurl = {https://ui.adsabs.harvard.edu/abs/2019PASJ...71..100S}
}

@ARTICLE{2023ApJ...957...99S,
       author = {{Suzuki}, Takeru K.},
        title = "{MHD in a Cylindrical Shearing Box. II. Intermittent Bursts and Substructures in MRI Turbulence}",
      journal = {\apj},
         year = 2023,
        month = nov,
       volume = {957},
       number = {2},
          eid = {99},
        pages = {99},
          doi = {10.3847/1538-4357/acfb88},
archivePrefix = {arXiv},
       eprint = {2305.12112},
 primaryClass = {astro-ph.HE},
       adsurl = {https://ui.adsabs.harvard.edu/abs/2023ApJ...957...99S}
}

@ARTICLE{2016A&A...596A..74S,
       author = {{Suzuki}, Takeru K. and {Ogihara}, Masahiro and {Morbidelli}, Alessandro and {Crida}, Aur{\'e}lien and {Guillot}, Tristan},
        title = "{Evolution of protoplanetary discs with magnetically driven disc winds}",
      journal = {\aap},
         year = 2016,
        month = dec,
       volume = {596},
          eid = {A74},
        pages = {A74},
          doi = {10.1051/0004-6361/201628955},
archivePrefix = {arXiv},
       eprint = {1609.00437},
 primaryClass = {astro-ph.EP},
       adsurl = {https://ui.adsabs.harvard.edu/abs/2016A&A...596A..74S}
}

@ARTICLE{2014ApJ...791..137B,
       author = {{Bai}, Xue-Ning},
        title = "{Hall-effect-Controlled Gas Dynamics in Protoplanetary Disks. I. Wind Solutions at the Inner Disk}",
      journal = {\apj},
         year = 2014,
        month = aug,
       volume = {791},
       number = {2},
          eid = {137},
        pages = {137},
          doi = {10.1088/0004-637X/791/2/137},
archivePrefix = {arXiv},
       eprint = {1402.7102},
 primaryClass = {astro-ph.EP},
       adsurl = {https://ui.adsabs.harvard.edu/abs/2014ApJ...791..137B}
}

@ARTICLE{2024PASJ...76..616I,
       author = {{Iwasaki}, Kazunari and {Tomida}, Kengo and {Takasao}, Shinsuke and {Okuzumi}, Satoshi and {Suzuki}, Takeru K.},
        title = "{Dynamics near the inner dead-zone edges in a proprotoplanetary disk}",
      journal = {\pasj},
         year = 2024,
        month = aug,
       volume = {76},
       number = {4},
        pages = {616-652},
          doi = {10.1093/pasj/psae036},
archivePrefix = {arXiv},
       eprint = {2401.03733},
 primaryClass = {astro-ph.EP},
       adsurl = {https://ui.adsabs.harvard.edu/abs/2024PASJ...76..616I}
}

@ARTICLE{1995ApJ...440..742H,
       author = {{Hawley}, John F. and {Gammie}, Charles F. and {Balbus}, Steven A.},
        title = "{Local Three-dimensional Magnetohydrodynamic Simulations of Accretion Disks}",
      journal = {\apj},
         year = 1995,
        month = feb,
       volume = {440},
        pages = {742},
          doi = {10.1086/175311},
       adsurl = {https://ui.adsabs.harvard.edu/abs/1995ApJ...440..742H}
}

@ARTICLE{1996ApJ...457..355G,
       author = {{Gammie}, Charles F.},
        title = "{Layered Accretion in T Tauri Disks}",
      journal = {\apj},
         year = 1996,
        month = jan,
       volume = {457},
        pages = {355},
          doi = {10.1086/176735},
       adsurl = {https://ui.adsabs.harvard.edu/abs/1996ApJ...457..355G}
}

@ARTICLE{2009ApJ...691L..49S,
       author = {{Suzuki}, Takeru K. and {Inutsuka}, Shu-ichiro},
        title = "{Disk Winds Driven by Magnetorotational Instability and Dispersal of Protoplanetary Disks}",
      journal = {\apjl},
         year = 2009,
        month = jan,
       volume = {691},
       number = {1},
        pages = {L49-L54},
          doi = {10.1088/0004-637X/691/1/L49},
archivePrefix = {arXiv},
       eprint = {0812.0844},
 primaryClass = {astro-ph},
       adsurl = {https://ui.adsabs.harvard.edu/abs/2009ApJ...691L..49S}
}

@ARTICLE{2016ApJ...817...52M,
       author = {{Mori}, Shoji and {Okuzumi}, Satoshi},
        title = "{Electron Heating in Magnetorotational Instability: Implications for Turbulence Strength in the Outer Regions of Protoplanetary Disks}",
      journal = {\apj},
         year = 2016,
        month = jan,
       volume = {817},
       number = {1},
          eid = {52},
        pages = {52},
          doi = {10.3847/0004-637X/817/1/52},
archivePrefix = {arXiv},
       eprint = {1505.04896},
 primaryClass = {astro-ph.EP},
       adsurl = {https://ui.adsabs.harvard.edu/abs/2016ApJ...817...52M}
}

@ARTICLE{2011ApJ...742...65O,
       author = {{Okuzumi}, Satoshi and {Hirose}, Shigenobu},
        title = "{Modeling Magnetorotational Turbulence in Protoplanetary Disks with Dead Zones}",
      journal = {\apj},
         year = 2011,
        month = dec,
       volume = {742},
       number = {2},
          eid = {65},
        pages = {65},
          doi = {10.1088/0004-637X/742/2/65},
archivePrefix = {arXiv},
       eprint = {1108.4892},
 primaryClass = {astro-ph.EP},
       adsurl = {https://ui.adsabs.harvard.edu/abs/2011ApJ...742...65O}
}

@ARTICLE{2018A&A...617A.117R,
       author = {{Riols}, A. and {Lesur}, G.},
        title = "{Dust settling and rings in the outer regions of protoplanetary discs subject to ambipolar diffusion}",
      journal = {\aap},
         year = 2018,
        month = sep,
       volume = {617},
          eid = {A117},
        pages = {A117},
          doi = {10.1051/0004-6361/201833212},
archivePrefix = {arXiv},
       eprint = {1805.00458},
 primaryClass = {astro-ph.EP},
       adsurl = {https://ui.adsabs.harvard.edu/abs/2018A&A...617A.117R}
}

@ARTICLE{2015MNRAS.454.1117S,
       author = {{Simon}, Jacob B. and {Lesur}, Geoffroy and {Kunz}, Matthew W. and {Armitage}, Philip J.},
        title = "{Magnetically driven accretion in protoplanetary discs}",
      journal = {\mnras},
         year = 2015,
        month = nov,
       volume = {454},
       number = {1},
        pages = {1117-1131},
          doi = {10.1093/mnras/stv2070},
archivePrefix = {arXiv},
       eprint = {1508.00904},
 primaryClass = {astro-ph.SR},
       adsurl = {https://ui.adsabs.harvard.edu/abs/2015MNRAS.454.1117S}
}

@ARTICLE{2011ApJ...736..144B,
       author = {{Bai}, Xue-Ning and {Stone}, James M.},
        title = "{Effect of Ambipolar Diffusion on the Nonlinear Evolution of Magnetorotational Instability in Weakly Ionized Disks}",
      journal = {\apj},
         year = 2011,
        month = aug,
       volume = {736},
       number = {2},
          eid = {144},
        pages = {144},
          doi = {10.1088/0004-637X/736/2/144},
archivePrefix = {arXiv},
       eprint = {1103.1380},
 primaryClass = {astro-ph.EP},
       adsurl = {https://ui.adsabs.harvard.edu/abs/2011ApJ...736..144B}
}

@ARTICLE{2000ApJ...530..464F,
       author = {{Fleming}, Timothy P. and {Stone}, James M. and {Hawley}, John F.},
        title = "{The Effect of Resistivity on the Nonlinear Stage of the Magnetorotational Instability in Accretion Disks}",
      journal = {\apj},
         year = 2000,
        month = feb,
       volume = {530},
       number = {1},
        pages = {464-477},
          doi = {10.1086/308338},
archivePrefix = {arXiv},
       eprint = {astro-ph/0001164},
 primaryClass = {astro-ph},
       adsurl = {https://ui.adsabs.harvard.edu/abs/2000ApJ...530..464F}
}

@ARTICLE{2002ApJ...581..988C,
       author = {{Casse}, Fabien and {Keppens}, Rony},
        title = "{Magnetized Accretion-Ejection Structures: 2.5-dimensional Magnetohydrodynamic Simulations of Continuous Ideal Jet Launching from Resistive Accretion Disks}",
      journal = {\apj},
         year = 2002,
        month = dec,
       volume = {581},
       number = {2},
        pages = {988-1001},
          doi = {10.1086/344340},
archivePrefix = {arXiv},
       eprint = {astro-ph/0208459},
 primaryClass = {astro-ph},
       adsurl = {https://ui.adsabs.harvard.edu/abs/2002ApJ...581..988C}
}

@ARTICLE{2013ApJ...778L..14A,
       author = {{Armitage}, Philip J. and {Simon}, Jacob B. and {Martin}, Rebecca G.},
        title = "{Two Timescale Dispersal of Magnetized Protoplanetary Disks}",
      journal = {\apjl},
         year = 2013,
        month = nov,
       volume = {778},
       number = {1},
          eid = {L14},
        pages = {L14},
          doi = {10.1088/2041-8205/778/1/L14},
archivePrefix = {arXiv},
       eprint = {1310.6745},
 primaryClass = {astro-ph.EP},
       adsurl = {https://ui.adsabs.harvard.edu/abs/2013ApJ...778L..14A}
}

@ARTICLE{2016ApJ...821...80B,
       author = {{Bai}, Xue-Ning},
        title = "{Towards a Global Evolutionary Model of Protoplanetary Disks}",
      journal = {\apj},
         year = 2016,
        month = apr,
       volume = {821},
       number = {2},
          eid = {80},
        pages = {80},
          doi = {10.3847/0004-637X/821/2/80},
archivePrefix = {arXiv},
       eprint = {1603.00484},
 primaryClass = {astro-ph.EP},
       adsurl = {https://ui.adsabs.harvard.edu/abs/2016ApJ...821...80B}
}

@ARTICLE{2017ApJ...845...31H,
       author = {{Hasegawa}, Yasuhiro and {Okuzumi}, Satoshi and {Flock}, Mario and {Turner}, Neal J.},
        title = "{Magnetically Induced Disk Winds and Transport in the HL Tau Disk}",
      journal = {\apj},
         year = 2017,
        month = aug,
       volume = {845},
       number = {1},
          eid = {31},
        pages = {31},
          doi = {10.3847/1538-4357/aa7d55},
archivePrefix = {arXiv},
       eprint = {1706.09565},
 primaryClass = {astro-ph.EP},
       adsurl = {https://ui.adsabs.harvard.edu/abs/2017ApJ...845...31H}
}

@ARTICLE{2019ApJ...879...98C,
       author = {{Chambers}, John},
        title = "{An Analytic Model for an Evolving Protoplanetary Disk with a Disk Wind}",
      journal = {\apj},
         year = 2019,
        month = jul,
       volume = {879},
       number = {2},
          eid = {98},
        pages = {98},
          doi = {10.3847/1538-4357/ab2537},
       adsurl = {https://ui.adsabs.harvard.edu/abs/2019ApJ...879...98C}
}

@ARTICLE{1981PThPS..70...35H,
       author = {{Hayashi}, C.},
        title = "{Structure of the Solar Nebula, Growth and Decay of Magnetic Fields and Effects of Magnetic and Turbulent Viscosities on the Nebula}",
      journal = {Progress of Theoretical Physics Supplement},
         year = 1981,
        month = jan,
       volume = {70},
        pages = {35-53},
          doi = {10.1143/PTPS.70.35},
       adsurl = {https://ui.adsabs.harvard.edu/abs/1981PThPS..70...35H}
}

@ARTICLE{2026JGRE..13109265S,
       author = {{Sato}, Masahiko and {Kimura}, Yuki and {Hatakeyama}, Tadahiro and {Nakamura}, Tomoki and {Okuzumi}, Satoshi and {Watanabe}, Sei-ichiro and {Sugita}, Seiji and {Tanaka}, Satoshi and {Tachibana}, Shogo and {Yurimoto}, Hisayoshi and {Noguchi}, Takaaki and {Okazaki}, Ryuji and {Yabuta}, Hikaru and {Naraoka}, Hiroshi and {Sakamoto}, Kanako and {Yada}, Toru and {Nishimura}, Masahiro and {Nakato}, Aiko and {Miyazaki}, Akiko and {Yogata}, Kasumi and {Abe}, Masanao and {Okada}, Tatsuaki and {Usui}, Tomohiro and {Yoshikawa}, Makoto and {Saiki}, Takanao and {Terui}, Fuyuto and {Nakazawa}, Satoru and {Tsuda}, Yuichi},
        title = "{Characteristics of Natural Remanence Records in Fine-Grained Particles Returned From Asteroid Ryugu}",
      journal = {Journal of Geophysical Research (Planets)},
         year = 2026,
        month = feb,
       volume = {131},
       number = {2},
          eid = {e2025JE009265},
        pages = {e2025JE009265},
          doi = {10.1029/2025JE009265},
archivePrefix = {arXiv},
       eprint = {2602.20806},
 primaryClass = {astro-ph.EP},
       adsurl = {https://ui.adsabs.harvard.edu/abs/2026JGRE..13109265S}
}

@ARTICLE{2011ApJ...735....8P,
       author = {{Perez-Becker}, Daniel and {Chiang}, Eugene},
        title = "{Surface Layer Accretion in Conventional and Transitional Disks Driven by Far-ultraviolet Ionization}",
      journal = {\apj},
         year = 2011,
        month = jul,
       volume = {735},
       number = {1},
          eid = {8},
        pages = {8},
          doi = {10.1088/0004-637X/735/1/8},
archivePrefix = {arXiv},
       eprint = {1104.2320},
 primaryClass = {astro-ph.EP},
       adsurl = {https://ui.adsabs.harvard.edu/abs/2011ApJ...735....8P}
}

@ARTICLE{2025ApJ...991L...6T,
       author = {{Teague}, Richard and {Lankhaar}, Boy and {Andrews}, Sean M. and {Qi}, Chunhua and {Fu}, Roger R. and {Wilner}, David J. and {Biersteker}, John B. and {Najita}, Joan R.},
        title = "{A Radially Resolved Magnetic Field Threading the Disk of TW Hya}",
      journal = {\apjl},
         year = 2025,
        month = sep,
       volume = {991},
       number = {1},
          eid = {L6},
        pages = {L6},
          doi = {10.3847/2041-8213/adff4d},
archivePrefix = {arXiv},
       eprint = {2509.09450},
 primaryClass = {astro-ph.EP},
       adsurl = {https://ui.adsabs.harvard.edu/abs/2025ApJ...991L...6T}
}

@ARTICLE{1996ApJ...463..656S,
       author = {{Stone}, James M. and {Hawley}, John F. and {Gammie}, Charles F. and {Balbus}, Steven A.},
        title = "{Three-dimensional Magnetohydrodynamical Simulations of Vertically Stratified Accretion Disks}",
      journal = {\apj},
         year = 1996,
        month = jun,
       volume = {463},
        pages = {656},
          doi = {10.1086/177280},
       adsurl = {https://ui.adsabs.harvard.edu/abs/1996ApJ...463..656S}
}

@ARTICLE{2025ApJ...992...85M,
       author = {{Mori}, Shoji and {Bai}, Xue-Ning and {Tomida}, Kengo},
        title = "{Radiative Nonideal Magnetohydrodynamic Simulations of Inner Protoplanetary Disks: Temperature Structures, Asymmetric Winds, and Episodic Surface Accretion}",
      journal = {\apj},
         year = 2025,
        month = oct,
       volume = {992},
       number = {1},
          eid = {85},
        pages = {85},
          doi = {10.3847/1538-4357/adf8d7},
archivePrefix = {arXiv},
       eprint = {2508.03624},
 primaryClass = {astro-ph.EP},
       adsurl = {https://ui.adsabs.harvard.edu/abs/2025ApJ...992...85M}
}

@ARTICLE{2012MNRAS.422.2737W,
       author = {{Wardle}, Mark and {Salmeron}, Raquel},
        title = "{Hall diffusion and the magnetorotational instability in protoplanetary discs}",
      journal = {\mnras},
         year = 2012,
        month = jun,
       volume = {422},
       number = {4},
        pages = {2737-2755},
          doi = {10.1111/j.1365-2966.2011.20022.x},
archivePrefix = {arXiv},
       eprint = {1103.3562},
 primaryClass = {astro-ph.EP},
       adsurl = {https://ui.adsabs.harvard.edu/abs/2012MNRAS.422.2737W}
}

@ARTICLE{2008MNRAS.385.1494K,
       author = {{Kunz}, Matthew W.},
        title = "{On the linear stability of weakly ionized, magnetized planar shear flows}",
      journal = {\mnras},
         year = 2008,
        month = apr,
       volume = {385},
       number = {3},
        pages = {1494-1510},
          doi = {10.1111/j.1365-2966.2008.12928.x},
archivePrefix = {arXiv},
       eprint = {0801.0974},
 primaryClass = {astro-ph},
       adsurl = {https://ui.adsabs.harvard.edu/abs/2008MNRAS.385.1494K}
}

@ARTICLE{2001ApJ...552..235B,
       author = {{Balbus}, Steven A. and {Terquem}, Caroline},
        title = "{Linear Analysis of the Hall Effect in Protostellar Disks}",
      journal = {\apj},
         year = 2001,
        month = may,
       volume = {552},
       number = {1},
        pages = {235-247},
          doi = {10.1086/320452},
archivePrefix = {arXiv},
       eprint = {astro-ph/0010229},
 primaryClass = {astro-ph},
       adsurl = {https://ui.adsabs.harvard.edu/abs/2001ApJ...552..235B}
}

@ARTICLE{1999MNRAS.307..849W,
       author = {{Wardle}, Mark},
        title = "{The Balbus-Hawley instability in weakly ionized discs}",
      journal = {\mnras},
         year = 1999,
        month = aug,
       volume = {307},
       number = {4},
        pages = {849-856},
          doi = {10.1046/j.1365-8711.1999.02670.x},
archivePrefix = {arXiv},
       eprint = {astro-ph/9809349},
 primaryClass = {astro-ph},
       adsurl = {https://ui.adsabs.harvard.edu/abs/1999MNRAS.307..849W}
}

@ARTICLE{2013ApJ...772...96B,
       author = {{Bai}, Xue-Ning},
        title = "{Wind-driven Accretion in Protoplanetary Disks. II. Radial Dependence and Global Picture}",
      journal = {\apj},
         year = 2013,
        month = aug,
       volume = {772},
       number = {2},
          eid = {96},
        pages = {96},
          doi = {10.1088/0004-637X/772/2/96},
archivePrefix = {arXiv},
       eprint = {1305.7232},
 primaryClass = {astro-ph.EP},
       adsurl = {https://ui.adsabs.harvard.edu/abs/2013ApJ...772...96B}
}

@ARTICLE{2024MNRAS.530.5131S,
       author = {{Sarafidou}, Eleftheria and {Gressel}, Oliver and {Picogna}, Giovanni and {Ercolano}, Barbara},
        title = "{Hall-magnetohydrodynamic simulations of X-ray photoevaporative protoplanetary disc winds}",
      journal = {\mnras},
         year = 2024,
        month = jun,
       volume = {530},
       number = {4},
        pages = {5131-5142},
          doi = {10.1093/mnras/stae1151},
archivePrefix = {arXiv},
       eprint = {2310.01985},
 primaryClass = {astro-ph.EP},
       adsurl = {https://ui.adsabs.harvard.edu/abs/2024MNRAS.530.5131S}
}

@ARTICLE{2024ApJ...972..128R,
       author = {{Rea}, David G. and {Simon}, Jacob B. and {Carrera}, Daniel and {Lesur}, Geoffroy and {Lyra}, Wladimir and {Sengupta}, Debanjan and {Yang}, Chao-Chin and {Youdin}, Andrew N.},
        title = "{Magnetically Driven Turbulence in the Inner Regions of Protoplanetary Disks}",
      journal = {\apj},
         year = 2024,
        month = sep,
       volume = {972},
       number = {1},
          eid = {128},
        pages = {128},
          doi = {10.3847/1538-4357/ad57c5},
archivePrefix = {arXiv},
       eprint = {2404.07265},
 primaryClass = {astro-ph.EP},
       adsurl = {https://ui.adsabs.harvard.edu/abs/2024ApJ...972..128R}
}

@ARTICLE{2005ApJ...628L..65F,
       author = {{Furlan}, E. and {Calvet}, N. and {D'Alessio}, P. and {Hartmann}, L. and {Forrest}, W.~J. and {Watson}, D.~M. and {Uchida}, K.~I. and {Sargent}, B. and {Green}, J.~D. and {Herter}, T.~L.},
        title = "{Colors of Classical T Tauri Stars in Taurus Derived from Spitzer Infrared Spectrograph Spectra: Indication of Dust Settling}",
      journal = {\apjl},
         year = 2005,
        month = jul,
       volume = {628},
       number = {1},
        pages = {L65-L68},
          doi = {10.1086/432540},
       adsurl = {https://ui.adsabs.harvard.edu/abs/2005ApJ...628L..65F}
}

@ARTICLE{2026arXiv260623815B,
       author = {{Bolchini}, Massimiliano and {Rosotti}, Giovanni and {Villenave}, Marion and {Garufi}, Antonio and {Benisty}, Myriam and {Birnstiel}, Tilman and {Facchini}, Stefano and {Testi}, Leonardo},
        title = "{Interpreting the scattering surface in protoplanetary disks}",
      journal = {arXiv e-prints},
         year = 2026,
        month = jun,
          eid = {arXiv:2606.23815},
        pages = {arXiv:2606.23815},
          doi = {10.48550/arXiv.2606.23815},
archivePrefix = {arXiv},
       eprint = {2606.23815},
 primaryClass = {astro-ph.EP},
       adsurl = {https://ui.adsabs.harvard.edu/abs/2026arXiv260623815B}
}
\end{document}